\documentclass[fleqn,usenatbib]{mnras}

\usepackage{newtxtext,newtxmath}
\usepackage[T1]{fontenc}

\DeclareRobustCommand{\VAN}[3]{#2}
\let\VANthebibliography\thebibliography
\def\thebibliography{\DeclareRobustCommand{\VAN}[3]{##3}\VANthebibliography}

\usepackage{graphicx}
\usepackage{placeins}
\usepackage{hyperref}
\usepackage{multirow}
\usepackage{float}
\usepackage{caption}
\usepackage{subcaption}
\usepackage{color}

\title[Velocity fields and WR-star spectra]{Exploring the influence of different velocity fields on Wolf-Rayet star spectra}

\author[R. R. Lefever et al.]{R.R. Lefever$^{1,2}$\thanks{E-mail: roel.lefever@uni-heidelberg.de},
          A.A.C. Sander$^{1}$,
          T. Shenar$^{2,3}$,
          L. G. Poniatowski$^{2}$,
          K. Dsilva$^{2,4}$
          and H. Todt$^{5}$
          \\
          $^1$Zentrum f\"ur Astronomie der Universit\"at Heidelberg, Astronomisches Rechen-Institut, M\"onchhofstr. 12-14, 69120 Heidelberg, Germany\\
          $^2$Institute for Astronomy (IvS), KU Leuven, Celestijnenlaan 200D, 3000 Leuven, Belgium\\
          $^3$Anton Pannekoek Institute for Astronomy, Science Park 904, 1098 XH, Amsterdam, The Netherlands\\
          $^4$Faculté des Sciences, Université Libre de Bruxelles, Boulevard du Triomphe, ACC.2, 1050 Bruxelles\\
          $^5$Institut für Physik und Astronomie, Universität Potsdam, Karl-Liebknecht-Str. 24/25, D-14476 Potsdam, Germany
          }

\date{Accepted 2023 February 24. Received 2023 February 24; in original form 2023 January 13}

\pubyear{2023}

\graphicspath{{./figures/}}

\begin{document}
\label{firstpage}
\pagerange{\pageref{firstpage}--\pageref{lastpage}}
\maketitle

\begin{abstract}
Given their strong stellar winds, Wolf-Rayet (WR) stars exhibit emission line spectra that are predominantly formed in
expanding atmospheric layers. 
The description of the wind velocity field $\varv(r)$ is therefore a crucial ingredient in the spectral analysis of WR stars, possibly influencing the determination of stellar parameters. 
In view of this, we perform a systematic study by simulating a sequence of WR-star spectra for different temperatures and mass-loss rates using $\beta$-type laws with $0.5\leq\beta\leq 20$. 
We quantify the impact of varying $\varv(r)$ by analysing diagnostic
lines and spectral classifications of emergent model spectra computed with the Potsdam Wolf-Rayet (PoWR) code. 
We additionally cross-check these models with hydrodynamically consistent --hydro-- model atmospheres. 
Our analysis confirms that the choice of the $\beta$-exponent has a strong impact on WR-star spectra, affecting line widths, line strengths and line profiles. 
In some parameter regimes, the entire range of WR subtypes could be covered. 
Comparison with observed WR stars and hydro models revealed that values of $\beta \gtrsim 8$ are unlikely to be realized in nature, but a range of $\beta$-values needs to be considered in spectral analysis.
UV spectroscopy is crucial here to avoid an underestimation of the terminal velocity $\varv_\infty$. 
Neither single- nor double-$\beta$ descriptions yield an acceptable approximation of the inner wind when compared to hydro models.
Instead, we find temperature shifts to lower $T_{2/3}$ when employing a hydro model.
Additionally, there are further hints that round-lined profiles seen in several early WN stars are an effect from non-$\beta$ velocity laws.
\end{abstract}


\section{Introduction}\label{sec:intro}

Massive stars (with $M_\ast \gtrsim 8\, M_\odot$) are known to have a large influence on the energetics of their host galaxies via ionizing radiation, strong stellar winds and end-of-life supernovae.
As a subset of massive stars, the massive Wolf-Rayet (WR) stars have exceptionally powerful stellar winds leading to typical mass loss rates of $\dot{M} = 10^{-5}\,\dots 10^{-4}\, M_\odot\,\mathrm{yr}^{-1}$ \citep[e.g.][]{crowther2007physical}, severely impacting their stellar evolution and enriching their circumstellar media.
The emission lines in this stellar wind are typically optically thick, acting as a strong driver for the stellar wind and causing a large outflow of material.
This material typically renders the hydrostatic surface of the WR star itself invisible and causes most, if not all, observed light to originate from the stellar wind of the star.
Due to the effect of line-driving, a profound understanding of the stellar wind is required to correctly derive the properties of a WR star.

The WR stars were first discovered and recognized by their peculiar spectra, containing a multitude of strong and broad emission lines, causing them to be assigned to a separate spectral class.
Based on the presence of certain elements, the WR star spectra are further divided into subclasses: the WN stars with strong nitrogen lines in the spectra, WC stars with prominent carbon lines and the WO stars with strong oxygen features.
WR stars are in most cases hydrogen-depleted or hydrogen-free, having lost their outer layers through continuous winds or eruptions \citep[e.g.][]{1983ApJ...274..302C, 1982MNRAS.201..451S, 2014ARA&A..52..487S} or binary mass-transfer \citep[e.g.][]{1967AcA....17..355P, 1998A&ARv...9...63V, 2014ApJ...789...10N, 2020A&A...634A..79S}.

About $\sim 90 \%$ of the known WR star population consists of massive stars that evolved past the main sequence and are typically core-helium burning \citep{2014A&A...565A..27H, 2019A&A...627A.151S}.
These evolved WR stars, called the classical WR stars, are thought to be progenitors of stellar-mass black holes \citep[e.g.][]{woosley2002}.
In contrast, there is a second group of very massive WR stars, displaying significant fractions of hydrogen in their spectra and still reside on, or close to, the main sequence. 
Thus, they are often considered to be core-hydrogen burning \citep{1997ApJ...477..792D}.

To infer the parameters of WR stars, sophisticated model atmospheres accounting for the expanding atmosphere are required, such as CMFGEN \citep[e.g.][]{1990A&A...231..116H, 1998ApJ...496..407H}, FASTWIND \citep[e.g.][]{1997A&A...323..488S}, or PoWR \citep[e.g.][]{grafener2002line, hamann2003temperature, sander2015consistent}.
These codes treat the wind in a non-LTE and chemically homogeneous framework, where the statistical rate equations and the full radiation transfer is computed instead of using the LTE approximations.
Given the numerical demands of the models, the velocity structure in the stellar wind is commonly pre-specified in the form of a so-called ``$\beta$-law''.
In the framework of line-driven winds, the $\beta$-law can be obtained from a solution of the (modified) CAK-theory (\citealt*{castor1975radiation}, \citealt{1986ApJ...311..701F}). 
It can be expressed as
\begin{equation}\label{eq:beta}
  \varv(r) = \varv_\infty \left(1 - \frac{R_\ast}{r}\right)^\beta,
\end{equation}
where $\varv_\infty$ denotes the terminal velocity of the wind, $R_\ast$ the stellar radius, and $r\in[R_\ast, +\infty)$ the radius in the stellar wind. 
The name-giving parameter $\beta$ is an exponent describing the steepness of velocity increase. 
In the case of O-type stars, \citet{1986A&A...164...86P} derived values of $\beta\approx 0.8$. 
All model codes mentioned above allow their user to pick their own value of $\beta$. 
Hence, the value is often used as another free parameter to improve the overall fit, e.g.\ by trying to reproduce the H$\alpha$-profile in B supergiants \citep[e.g.][]{Searle+2008}. 
For WR stars, the value of $\beta=1$ is commonly adopted \citep[e.g.][]{Hillier1988,Hamann+1988,HillierMiller1999}, and is also standard in pre-calculated model grids \citep[e.g.][]{hamann2004grids,sander2012galactic,todt2015potsdam}.
However, this assumption has also been challenged. 
Spectral modeling of observations carried out by \citet{robert1994optical} resulted in better fits of WR-star spectra for velocity laws with $\beta>3$. 
Studies by \citet{lepine1999wind} suggest values of $\beta\sim 5\text{ - } 20$, using the temporal evolution of line profile variations interpreted as clumping in the stellar wind. 
The latter values probe mostly the outer regions of the stellar wind, where the continuum is optically thin. 
Thus, one could argue that these high values might only apply to this part of the wind and a lower $\beta$ could be necessary at smaller radii. 
Interestingly, the analysis of the sonic point by \citet{2002A&A...389..162N}, coming from a completely different angle, also favours $\beta$ values of $5$ and larger.

Instead of pre-specifying a velocity law in WR-star models \citep[e.g.][]{graefener2008}, studies by \citet{grafener2005hydrodynamic}, \citet{sander2020nature} and \citet{sander2020driving} have constructed wind models where the velocity field is computed from solving the hydrodynamic equation consistently with the radiative transfer. 
However, hydrodynamically consistent studies of classical WR stars are computationally expensive and their resulting velocity fields cannot be easily parametrized \citep{grafener2005hydrodynamic,Sander2015proceedings,sander2020driving}. 
From the modelling of an early-type WC star, \citet{grafener2005hydrodynamic} suggested a
double-$\beta$-law 
\begin{equation}
  \label{eq:doublebeta}
    \varv(r) = \varv_\infty\left[\left(1-q\right) \left(1 - \frac{R_\ast}{r}\right)^{\beta_1} + q \left(1 - \frac{R_\ast}{r}\right)^{\beta_2}\right]
\end{equation}
\citep[as proposed in][]{HillierMiller1999} with $\beta_1 = 1$ and $\beta_2 = 50$ as an approximation of the hydrodynamic result.
The parameter $q$ is a weighting factor between 0 and 1.
This could qualitatively still be in line with the results of \citet{lepine1999wind}. 
This double-$\beta$ law is used to apply two separate $\beta$ laws to the inner wind and outer wind regime separately. 
As most of the observable line spectrum of a WR star is formed in its outer wind, a full parameter study of the double-$\beta$ law is unnecessary as the emergent spectral imprint of the inner $\beta$ law will be minimal.

Due to its reference radius $R_\ast$, Eq.\,\eqref{eq:beta} is connected to the stellar radius of a WR star. 
From Eq.\,\eqref{eq:beta}, we would expect the velocity to be zero there, meaning that $R_\ast$ should correspond to a hydrostatic radius. 
Typically, the stellar surface radius is defined at the point where the Rosseland mean optical depth $\tau_\mathrm{Ross}\approx 2/3$. 
Because of the optically thick winds of WR stars, this point -- which we denote as $R_{2/3}$ with a corresponding effective temperature $T_{2/3}$ -- is usually far out in the wind of the star, where we have a significant bulk movement away from the actual star and the hydrostatic equation is significantly violated. 
Instead, subsonic velocities are only reached much further inside at $\tau_\mathrm{Ross}\approx 20$ \citep[e.g.][]{2002A&A...389..162N,sander2020nature} although some hydrodynamical models also yield values of $\tau_\mathrm{Ross}\approx 5$ \citep[e.g.][]{grafener2005hydrodynamic}. 
To be safely in the hydrostatic regime, the inner boundary of PoWR models, e.g. in the published WR grids \citep{hamann2004grids,todt2015potsdam} is traditionally set to $\tau_\mathrm{Ross,cont} = 20$. 
The corresponding radius to this (continuum) optical depth is termed as $R_\ast$ with a formally defined effective temperature $T_\ast$ following from the Stefan-Boltzmann equation. 
Hence, for PoWR models, $T_\ast$ is typically an input parameter, while $T_{2/3}$ is an output of a converged atmosphere model.
\begin{figure}
	\centering
    \includegraphics[width=\hsize]{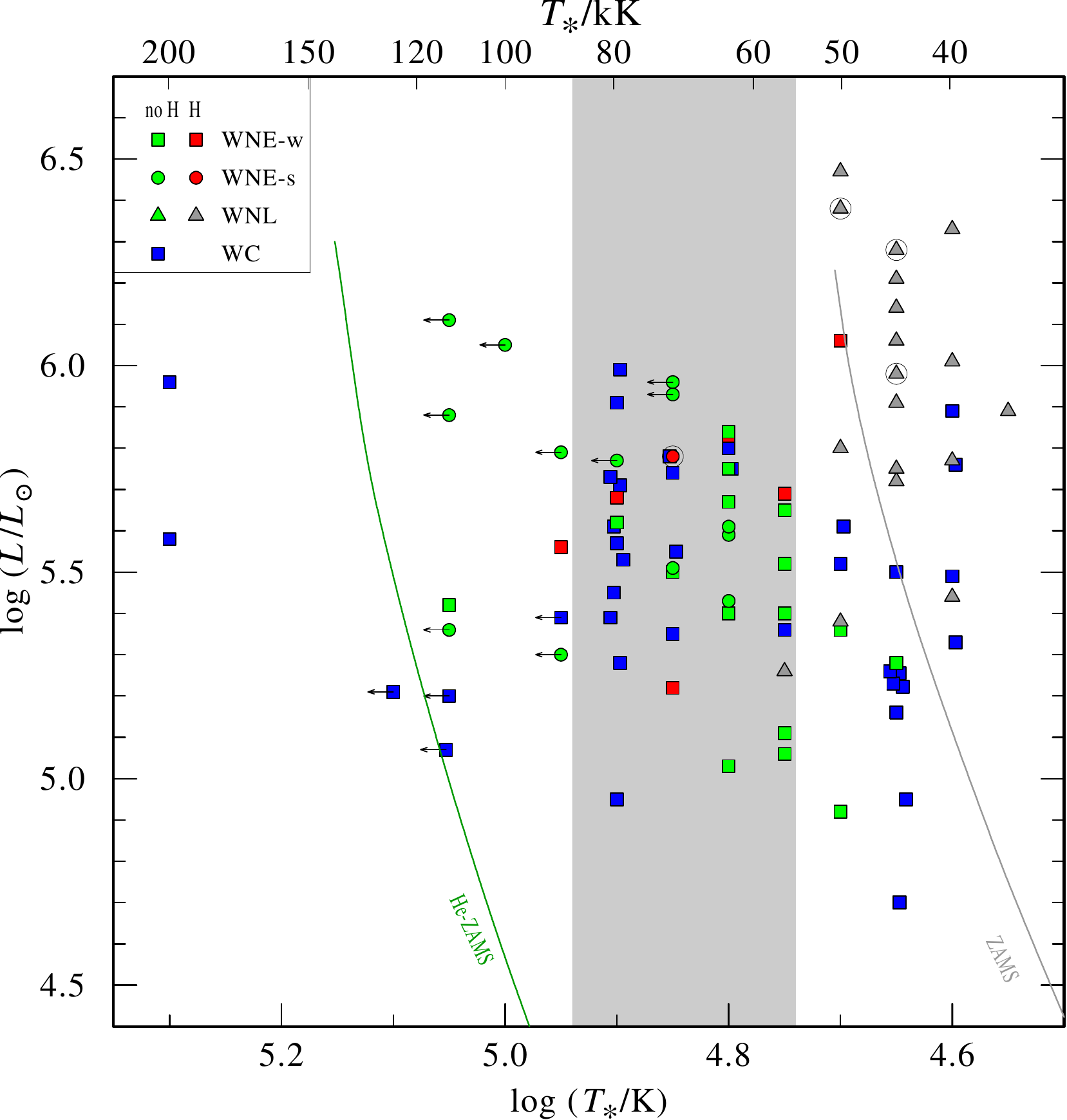}    
	\caption{Positions in the HR-diagram of a set of Galactic Wolf-Rayet stars, adapted from \citet{hamann2019galactic}. 
    The stellar temperatures $T_\ast$ are at $\tau_\mathrm{Ross,cont} = 20$.
    The shaded area denotes the temperatures used later on in the model sequences, see e.g. Table \ref{tab:gridmodel_params}.}
	\label{fig:comp}
\end{figure}
\begin{figure}
	\centering
    \includegraphics[width=\hsize]{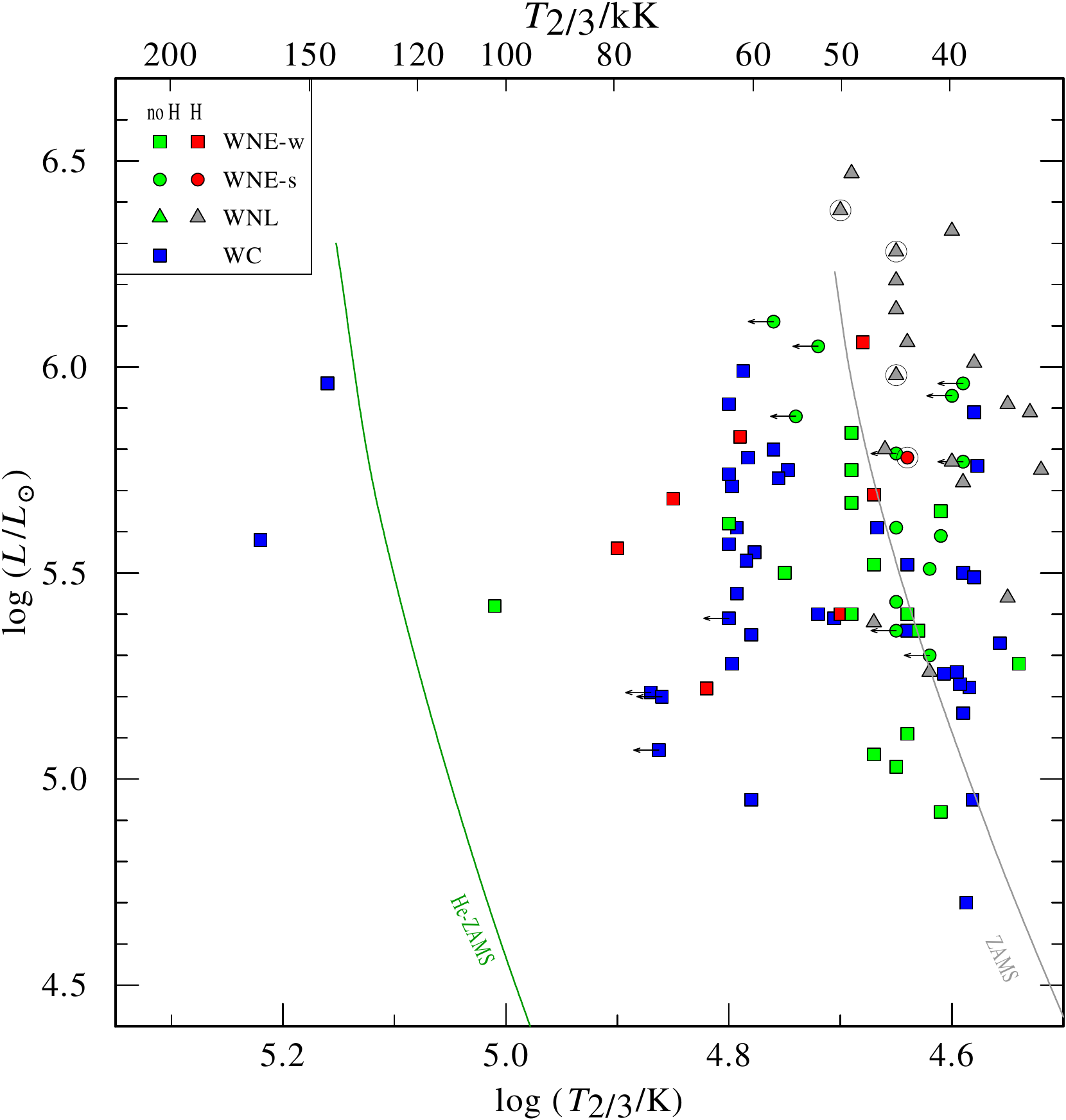}\caption{Same Galactic WR-stars as in Fig. \ref{fig:comp}, but now shown with $T_{2/3}$, the effective temperatures at $\tau_\mathrm{Ross}\approx 2/3$.}
	\label{fig:comp_T23}
\end{figure}
In stellar structure models, the effect of optically thick winds cannot be taken into account accurately as this would require a hydrodynamic treatment and the computation of the flux-weighted opacity, which deviates significantly from the Rosseland opacity in an expanding non-LTE environment \citep[illustrated e.g.\ in][]{sander2020driving,sander2022driving}.
This makes it difficult to empirically compare derived temperatures $T_\ast$ for WR stars in the Hertzsprung-Russell diagram (HRD). 
Some evolution codes provide an estimate of the wind correction, but these can be severely off compared to actual atmosphere calculations as shown by \citet{groh2014evolution}. 
However, also the usage of $T_\ast$ raises questions as indicated in Fig.\,\ref{fig:comp}. 
For He-burning classical WR (cWR) stars, the large spread in temperature is not expected, especially if the stars are almost or completely depleted in hydrogen \citep[e.g.][]{ekstrom2012grids}. 

Rather, these H-stripped stars should be located close to the helium zero-age main sequence (He-ZAMS). 
This discrepancy between observations and structure models is referred to as the \emph{Wolf-Rayet radius problem} \citep[e.g.][]{2018A&A...614A..86G}. 
As we will show later on in this work, the $T_{2/3}$-temperatures, which can be directly inferred from spectral analyses, are less affected by uncertainties in $R_\ast$. 
However, they do not reflect the hydrostatic layers of the stars. 
This is illustrated in the HRD from \citet{hamann2019galactic} now with $T_{2/3}$ instead of $T_\ast$ in Fig.\,\ref{fig:comp_T23}. 
In the $T_{2/3}$-HRD, almost no star of the WR population is located on the He-ZAMS, even for the completely hydrogen-free stars, emphasizing that these ``observable'' $T_{2/3}$-temperatures only reflect the extended wind layers and thus are problematic for any general comparison with evolutionary tracks. 

Two major solutions are suggested: either the actual hydrostatic radii from where the winds are launched are significantly smaller than derived from the empirical models using $\beta=1$ velocity laws, such as obtained in the prototypical calculations of hydrodynamically consistent atmosphere models \citep{grafener2005hydrodynamic,sander2020driving,sander2020nature}, or rather the hydrostatic radii are inflated \citep[e.g.][]{petrovic2006luminous, grafener2012stellar} and the winds are actually launched at approximately the radii determined from the empirical models. 
In fact, both phenomena might exist in nature with a branching depending on the precise stellar parameters \citet{2018A&A...614A..86G, poniatowski2021dynamically}. 

In this work, we will investigate how much the empirically derived results are affected by the common choice of $\beta = 1$ for the velocity law. 
Moreover, we will investigate whether the use of different $\beta$-type velocity laws could bring the empirical values of $T_\ast$ closer to the predictions from stellar structure theory. 
Following up on the early suggestion by \citet{hillier1991theory} that the $\beta$-law may result in ambiguous parameters such as the stellar radius $R_\ast$, we probe the impact of different $\beta$-values for a range of mass-loss rates and temperatures. 
\citet{hillier2003advances} concluded that the $\beta$ parameter did not have a significant impact on the line spectrum when varying $\beta$ between $1$ and $3$, but as we will show, this does not hold when considering larger $\beta$ values. 
Moreover, we compare the $\beta$-model efforts to trends from hydrodynamically consistent models and perform an exemplary approximation of a hydrodynamically consistent solution to gain first insights on the wind acceleration, $T_{2/3}$ and spectral appearance, preparing us for more tailored investigations of individual targets with hydrodynamically consistent models in the future.

The paper is organised as follows: the methodology and modelling used to conduct this study is explained in Sect.\,\ref{sec:methods}.
In Sect.\,\ref{sec:results}, we present the comparison and classification of the emergent spectra from our $\beta$-law models and further discuss the outcomes and implications in Sect.\,\ref{sec:discussion}. 
Afterwards, we compare our findings with results from hydrodynamically consistent modelling in Sect.\,\ref{sec:hydrocmp}.
A summary of our work, along with future prospects, is given in Sect.\,\ref{sec:conclusions}.

\section{Methods}\label{sec:methods}

The strong coupling between radiation field and population numbers cause the modelling of WR-star winds to be numerically demanding as the local thermodynamic equilibrium (LTE) approximation cannot be realistically assumed.
The Potsdam Wolf-Rayet (PoWR) code \citep{grafener2002line, hamann2003temperature, sander2015consistent} is one of the few codes capable of modelling these stellar winds in a non-LTE setting in the co-moving frame, assuming a spherically symmetric, stationary outflow and chemical homogeneity.
Unless the hydrodynamic branch is used \citep{Sander+2017}, the wind is described by the mass loss rate $\dot{M}$ and a pre-specified velocity field $\varv(r)$ with the help of a $\beta$-law with parameters $\beta$ and $\varv_\infty$.
The code computes the atmospheric stratification and emergent model spectra for a given set of stellar parameters $(M_\ast, T_\ast, R_\ast, L_\ast, \dot{M}, \varv_\infty, \beta)$. The emergent spectra are computed in the observer's frame such that they can be properly compared to observed spectra.

To study the impact of different values of $\beta$, we calculated several model sequences employing a range of $T_\ast$ values that maps the bulk of the empirically analysed Galactic WR stars \citep{hamann2019galactic,sander2019galactic}. 
The corresponding area in the Hertzsprung–Russell diagram (HRD) is illustrated in Fig.\,\ref{fig:comp}. 

For a fixed velocity law, WR wind models can be characterised by two main values: the stellar temperature $T_\ast$ and the \textit{transformed radius} $R_\mathrm{t}$ \citep{schmutz1989spectral}:
\begin{equation}\label{eq:transformed_radius}
    R_\text{t} = R_\ast \left[\frac{\varv_\infty}{2500\, \mathrm{km s}^{-1}}\Bigg/\frac{\sqrt{D} \dot{M}}{10^{-4}M_\odot\, \mathrm{yr}^{-1}}\right]^{2/3},
\end{equation}
\noindent with $D$ denoting the clumping factor \citep{1998A&A...335.1003H}. 
While having the dimension of a radius, the absolute value of $R_\text{t}$ has no direct meaning in the sense of a physically relevant radius. 
Instead, $R_\text{t}$ represents an emission measure introduced by \citep{schmutz1989spectral}, who discovered that line strengths in stellar models with the same $T_\ast$ are approximately equal for similar values of $R_\mathrm{t}$ \citep{schmutz1989spectral}, regardless of the individual values entering Eq.\eqref{eq:transformed_radius}. 
This allows for a scaling of the models with regard to the luminosity of the object.

Instead of $R_\mathrm{t}$ one can also define the \textit{transformed mass-loss rate},
\begin{equation}\label{eq:transformed_massloss}
    \dot{M}_\mathrm{t} = \dot{M} \sqrt{D} \left(\frac{1000\text{ km s$^{-1}$}}{\varv_\infty}\right)\left(\frac{10^6\ L_\odot}{L_\ast}\right)^{3/4},
\end{equation}
\noindent introduced by \citet{grafener2013stellar}, which denotes the mass-loss rate the star would have if it had no clumping ($D=1$), $\varv_\infty = 1000$ km\,s$^{-1}$ and $L_\ast = 10^6\, L_\odot$. 
In a similar fashion as with $R_\mathrm{t}$, also $\dot{M}_\mathrm{t}$ can be seen as a measure for the general emission line strength \citep[see, e.g,][for details]{sander2020nature}.
With the unit of $\dot{M}_\mathrm{t}$ being slightly more intuitive, we will denote our WR-star models via their respective $T_\ast$ and $\dot{M}_\mathrm{t}$ values: letters A, B and C denote $T_\ast$, while numbers 1,2 and 3 denote $\dot{M}_\mathrm{t}$, see e.g.\ Table \ref{tab:gridmodel_params}, with higher numbers corresponding to a higher $\dot{M}_\mathrm{t}$.

For each of our selected $T_\ast$-values, we pick the same three values of $\dot{M}_\mathrm{t}$. 
In our selection, we deliberately avoid the regime of very dense winds where we already know that $T_\ast$ becomes meaningless as the whole spectrum is generated in layers where $\varv \approx v_\infty$ \citep{hamann2004grids, hamann2006galactic}. 

We investigate three different types of WR models in this work: hydrogen-free WN stars, hydrogen-rich WNh stars, and carbon-rich WC stars.
The first type of models are resembling cWR stars with strong nitrogen lines in their spectra and are for convenience referred to as of ``WN type'' in the rest of this work. 
The WNh-star models have a significant amount of hydrogen at their surface ($X_\mathrm{H} = 0.5$) and are denoted as WNh models. 
The carbon-rich WC models represent the hydrogen-free WC stars with carbon and oxygen surface mass fractions of $X_\text{C} = 0.4$ and $X_\text{O} = 0.05$. 
These are denoted as WC models. 
\begin{table*}
    \caption{Stellar parameters of the models used in this study. 
    The distinction is made between models of WN stars (containing nitrogen, no hydrogen), WNh stars (containing nitrogen, presence of hydrogen with $X_H = 50\%$) and WC stars (carbon-rich and no hydrogen). 
    The stellar parameters are mainly equal for WN, WNh and WC-star models with the exception of $\log(\dot{M})$ and $\varv_\infty$.
    All models assume a luminosity of $\log(L_\ast/L_\odot) = 5.3$.}
    \centering
    \def\arraystretch{1.3}
    \begin{tabular}{|c|c|ccccccccc|}
        \hline
        \multicolumn{2}{|c|}{Model indices:} & A1 & A2 & A3 & B1 & B2 & B3 & C1 & C2 & C3\\
        \hline
        \multicolumn{2}{|c|}{$T_\ast\, [\mathrm{kK}]$} & 56.2 & 56.2 & 56.2 & 70.8 & 70.8 & 70.8 & 89.1 & 89.1 & 89.1 \\
        \multicolumn{2}{|c|}{$\log(\dot{M_\mathrm{t}}\, [M_\odot/\mathrm{year}])^a$} & -4.66 & -4.36 & -4.06 & -4.66 & -4.36 & -4.06 & -4.66 & -4.36 & -4.06\\
        \multicolumn{2}{|c|}{$R_\ast\, [R_\odot]$} & 4.72 & 4.72 & 4.72 & 2.98 & 2.98 & 2.98 & 1.88 & 1.88 & 1.88 \\
        \multicolumn{2}{|c|}{$R_\mathrm{t}\, [R_\odot]^b$} & 15.8 & 10.0 & 6.3 & 10.0 & 6.3 & 4.0 & 6.3 & 3.98 & 2.51\\
        \multirow{3}{*}{\shortstack[2]{$\log(\dot{M}\, [M_\odot/\mathrm{year}])$}} & WN, $\varv_\infty=1600$ km s$^{-1}$ & -5.28 & -4.98 & -4.68 & -5.28 & -4.98 & -4.68 & -5.28 & -4.98 & -4.68\\
         & WNh, $\varv_\infty=1000$ km s$^{-1}$ & -5.49 & -5.19 & -4.89 & -5.49 & -5.19 & -4.89 & -5.49 & -5.19 & -4.89\\
         & WC, $\varv_\infty=2000$ km s$^{-1}$ & -5.39 & -5.09 & -4.79 & -5.39 & -5.09 & -4.79 & -5.39 & -5.09 & -4.79\\
        \hline
        \multicolumn{11}{l}{$^a$ Transformed mass-loss rate, the mass-loss rate the star would have if $D=1$, $\varv_\infty = 1000$ km\,s$^{-1}$ and $L_\ast = 10^6\, L_\odot$.  see Eq.\,\eqref{eq:transformed_massloss}.} \\
        \multicolumn{11}{l}{$^b$ see Eq.\,\eqref{eq:transformed_radius}.}
    \end{tabular}
    \label{tab:gridmodel_params}
\end{table*}
For each of the three WR-types and each of the nine $(T_\ast,\dot{M}_\mathrm{t})$-combinations we calculate a sequence of models using $\beta$-velocity fields with $\beta = 0.5, 1, 2, 4, 8 \text{ and } 20$, leading to a total of 27 model sequences or 262 models.
An overview of their specific parameters is shown in Table \ref{tab:gridmodel_params}.
For the remainder of this paper, we will denote a model which uses the $\beta$-velocity law with $\beta=x$ as a $\beta x$-model. 
Hence, if $\beta=1$, we denote it as a $\beta1$-model, for $\beta=4$ we denote it as a $\beta4$-model, and so forth.
\begin{table*}
    \caption{Diagnostic spectral lines for WN stars \citep{smith1996three} and for WC stars \citep{crowther1998quantitative}. 
    The line strengths are measured by the peak-to-continuum ratios rather than the equivalent widths of the lines.}
    \centering
    \vspace{2mm}
    \def\arraystretch{1.3} % increase column height
    \begin{tabular}{|c|lllllll|}
        \hline
        class & \multicolumn{7}{|c|}{Element - ionisation stage - wavelength [\AA]} \\
        \hline
        WN$^a$: & \ion{He}{i} $\lambda$ 5875 & \ion{He}{ii} $\lambda$ 5411 & \ion{C}{iv} $\lambda$ 5808 & \ion{N}{iii} $\lambda$ 4640 & \ion{N}{iv} $\lambda$ 4057 & \ion{N}{v} $\lambda$ 4604 & \\
        WC: & \ion{He}{i} $\lambda$ 5875 & \ion{He}{ii} $\lambda$ 4686 & \ion{C}{ii} $\lambda$ 4267 & \ion{C}{iii} $\lambda$ 5696 & \ion{C}{iv} $\lambda$ 5808 & \ion{O}{v} $\lambda$ 5590 & \ion{O}{vi} $\lambda$ 3818 \\
        \hline
        \multicolumn{8}{l}{$^a$The same diagnostic lines were used for the WNh-star models.}
    \end{tabular}
    \label{tab:diagnostic_lines}
\end{table*}
To quantify the differences between the resulting model spectra, we employ the spectral classification criteria and diagnostics from \citet{smith1996three} for WN stars and from \citet{crowther1998quantitative} for WC stars. 
An overview of the diagnostic spectral lines used in this work is given in Table \ref{tab:diagnostic_lines}. 
These lines are especially sensitive to changes in stellar parameters.
Both \citet{smith1996three} and \citet{crowther1998quantitative} preferably use peak-to-continuum values -- the spectral line peaks in continuum units subtracted by 1 -- of spectral lines as indicators for line strengths and not the equivalent widths for emission lines.
This is due to the often blended emission lines in WR stars, caused by heavy broadening of the lines due to large Doppler shifts in the stellar wind.
Using the normalized spectra of our models, we automatize the classification scheme accounting for the broadening of spectral lines and the often accompanying line-blending. 
Based on the line strength ratios, the spectra are classified from the early to late subclasses WN2 to WN9 or WC4 to WC11.
For the comparison with hydrodynamically-consistent models, we make use of models calculated in the framework of \citet{sander2020nature}.

\section{Results from the $\beta$-sequences}\label{sec:results}

\begin{figure}
    \centering
    \includegraphics[width=\hsize]{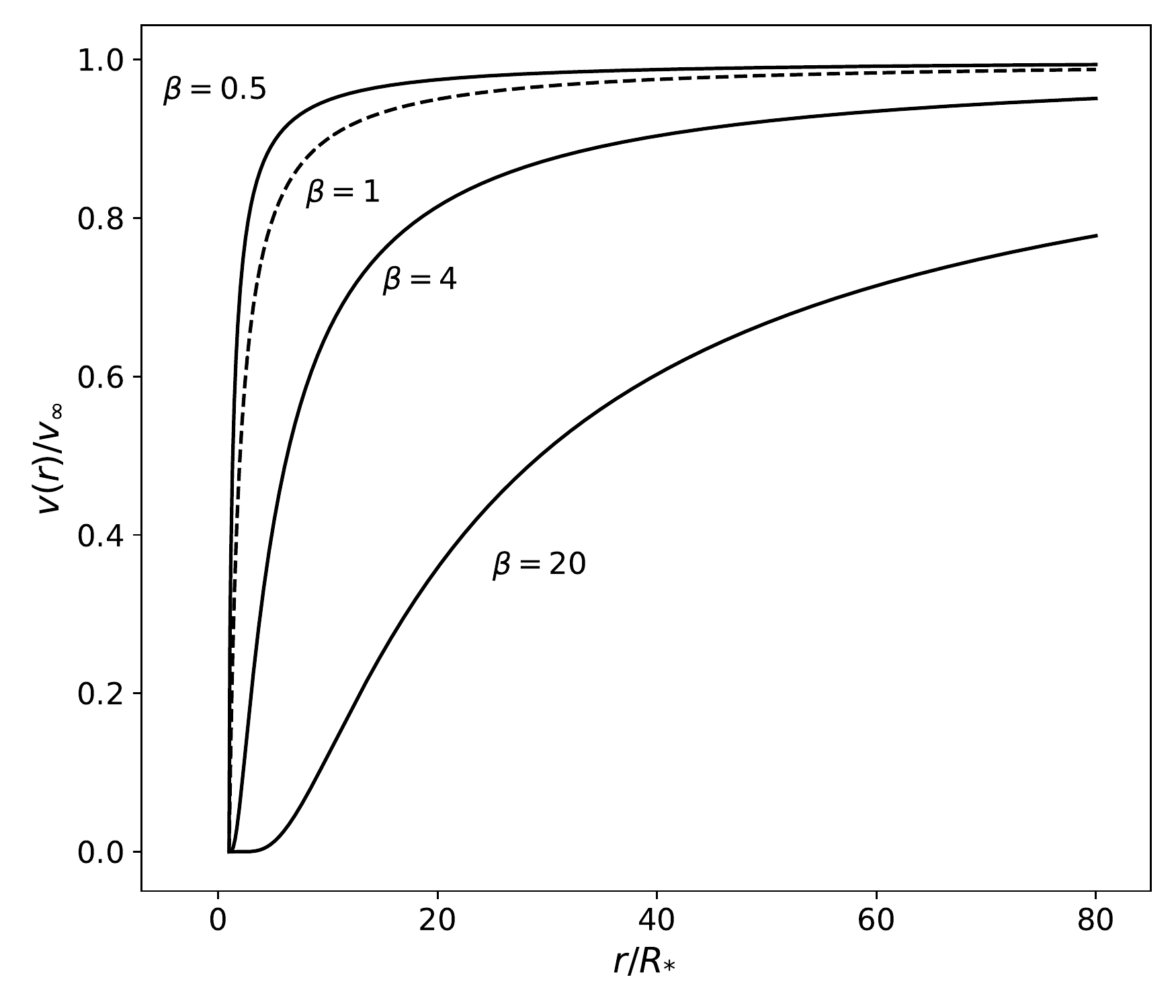}
    \caption{The $\beta$-velocity law for different values of $\beta$, starting at radius $r = R_\ast$. 
    The increase in velocity grows shallower for increasing values of $\beta$, reaching $\varv_\infty$ at large $r >> R_\ast$ in the stellar wind.
    The dashed line represents the $\beta1$-law as reference.}
    \label{fig:betalaws}
\end{figure}

An overview of the different $\beta$-velocity laws from our model sequences is given in Fig.\,\ref{fig:betalaws} with the $\beta1$-velocity law highlighted. 
The velocity fields in Fig.\,\ref{fig:betalaws} are normalized to their (terminal) value at infinity.
As clearly evident from the figure, the $\beta$-parameter from Eq.\,\eqref{eq:beta} reflects the steepness of the velocity field, meaning that with higher values of $\beta$ a given velocity is only reached further out in the wind.

\subsection{Line strengths}\label{subsec:strengths}

\begin{figure}
    \centering
    \includegraphics[width=\hsize]{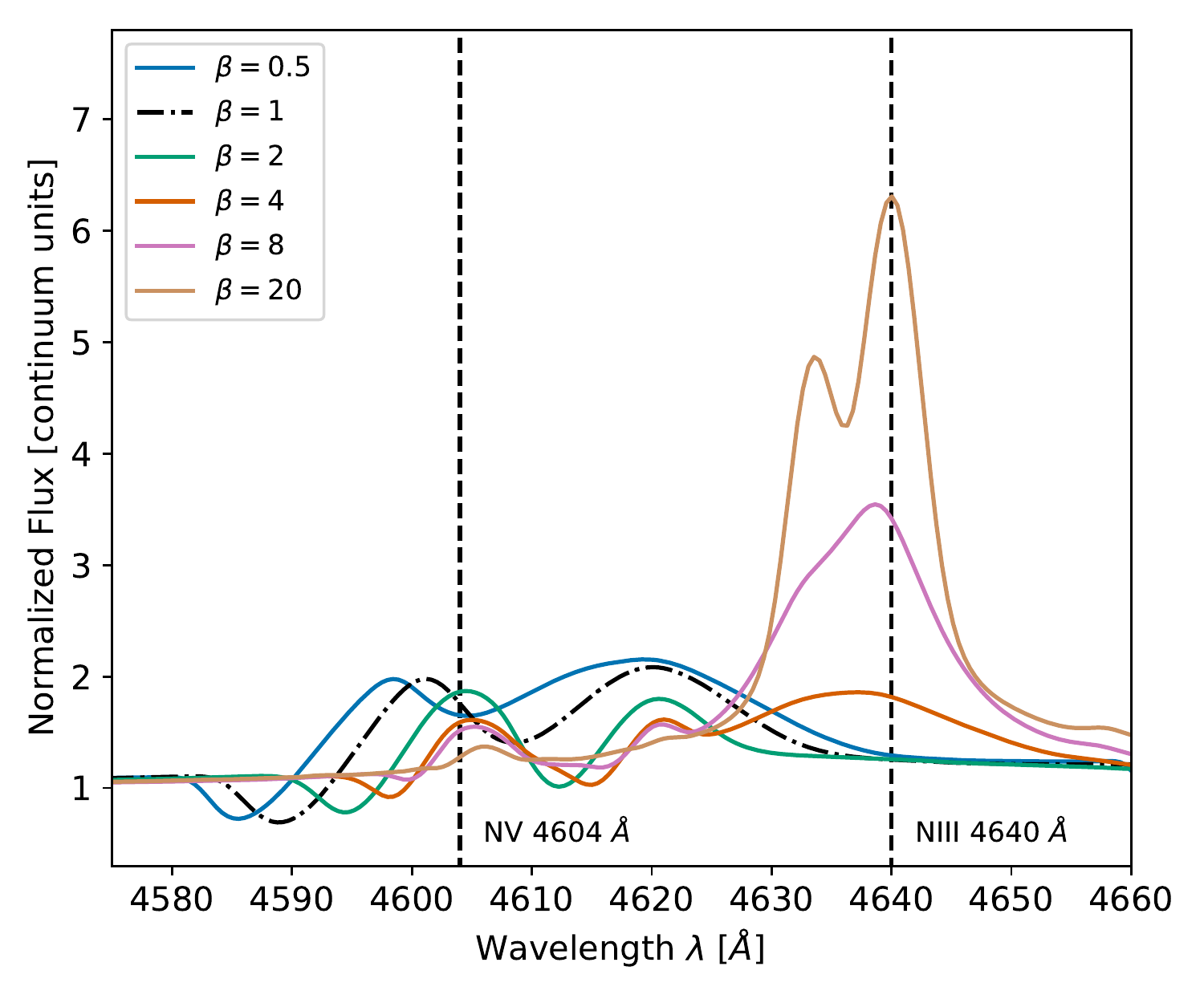}
    \caption{The diagnostic line pair \ion{N}{iii}-\textsc{v} $\lambda$ 4640-04 of the B2 WN-star model spectrum shown for different $\beta$-values. 
    Note the change in line strength ratio: for lower $\beta$ the \ion{N}{v}-line dominates, while for higher $\beta$ the \ion{N}{iii}-line dominates.
    The dash-dotted $\beta1$-model is shown as reference.} 
    \label{fig:wne_linewidths_comp}
\end{figure}

\begin{figure}
    \centering
    \includegraphics[width=\hsize]{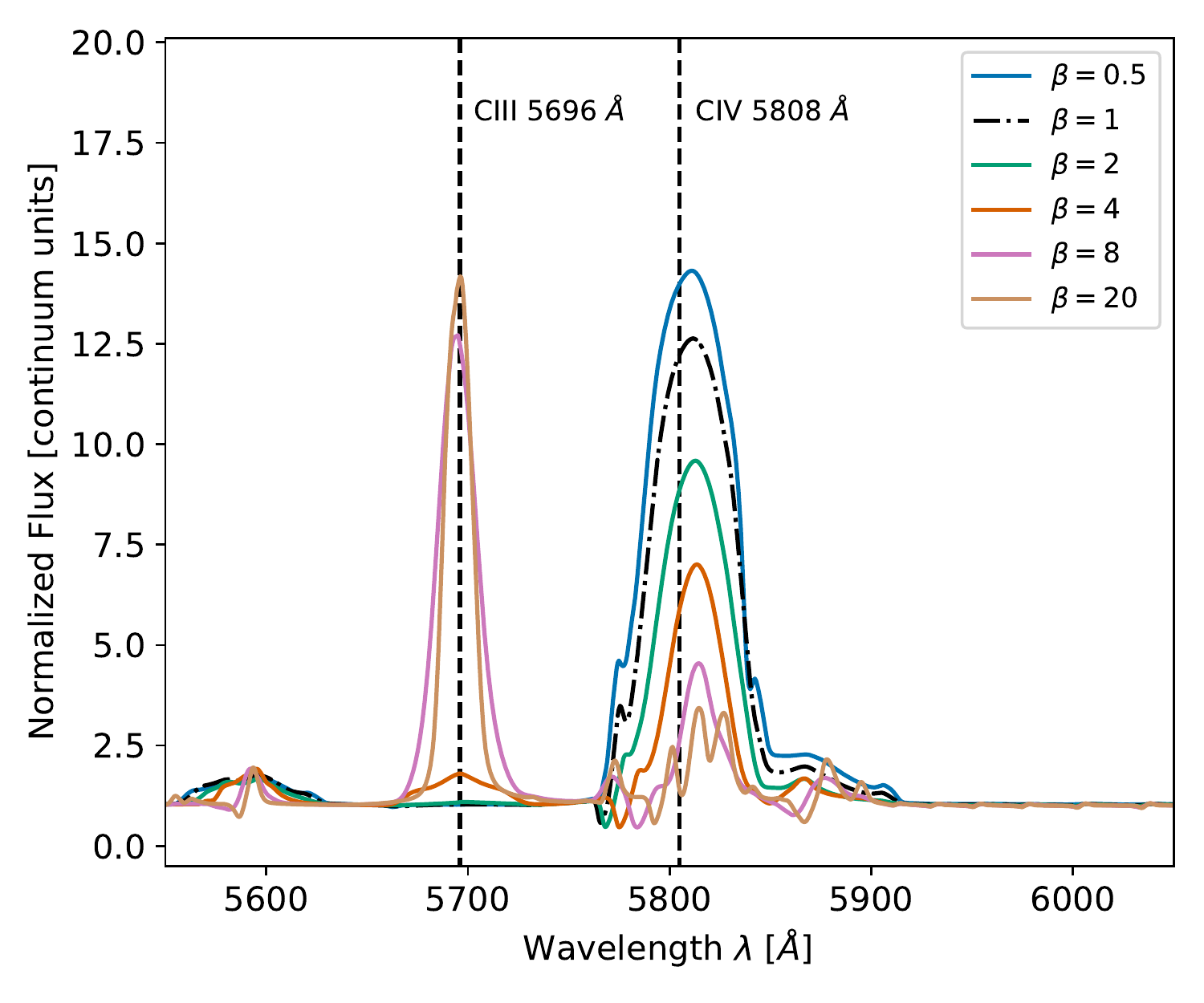}
    \caption{Similar to Fig.\,\ref{fig:wne_linewidths_comp}, but for the diagnostic line pair \ion{C}{iii}-\textsc{iv} $\lambda$ 5696-808 of the B2 WC-star model.
    The $\beta$-value changes the line strength ratio with the \ion{C}{iv}-line dominating for lower $\beta$ while the \ion{C}{iii}-line dominates for higher $\beta$.
    The $\beta1$-model is outlined in a dash-dotted style as a reference.}
    \label{fig:wc_linewidths_comp}
\end{figure}

To convey the influence of the $\beta$-parameter, we compare the emergent spectra of our atmosphere models along each $\beta$-sequence. 
With this term we denote the sequences of models that are identical in their stellar parameters and only differ in the $\beta$-value applied for their wind description. 
The influence of $\beta$ is most obvious when studying the diagnostic lines. An example is shown in Fig.\,\ref{fig:wne_linewidths_comp} where we plot the spectra of the WN B2 models (see Table \ref{tab:gridmodel_params} for the corresponding stellar parameters) in the \ion{N}{iii}-\textsc{v} $\lambda\, 4640 \text{-} 04$ \AA\, line region. 
The change in line strengths (peak-to-continuum ratios) is immediately apparent.
In the $\beta0.5 \text{ and } \beta1$ cases, the \ion{N}{iii} $\lambda\,4640$ \AA\, line is too weak to be distinguished from the continuum and the \ion{N}{v} $\lambda\,4604$ \AA\, line dominates.
While emergent spectra with $\beta2 \text{ and } \beta4$ laws show the \ion{N}{iii} and \ion{N}{v} lines being of comparable strength, the \ion{N}{iii} line dominates while the \ion{N}{v} line diminishes for $\beta8 \text{ and } \beta20$ laws.
This implies a strong decrease in the $\text{\ion{N}{v}}\,\lambda\, 4604\, \big/\, \text{\ion{N}{iii}}\,\lambda\, 4640$ line-strength ratio with increasing $\beta$.
While the changes in line strength for low $\beta$ values of 0.5 to 2 are small -- similarly to \citet{hillier1991theory} -- and may fall below the detection limit in real observations, more extreme $\beta$  values (4 to 20) lead to very significant changes.
The remaining diagnostic lines and accompanying ratios are affected in a similar way.
Analogous to the WN-star models, the \ion{C}{iii} $\lambda\,5696$ and \ion{C}{iv} $\lambda\,5808$ lines are affected in WC stars. We illustrate this in Fig.\,\ref{fig:wc_linewidths_comp}, where we present the WC B2 models with different $\beta$-values, highlighting that the \ion{C}{iii} line increases in strength while the \ion{C}{iv} line diminishes when increasing $\beta$.
This causes the $\text{\ion{C}{iv}}\,\lambda\, 5808\, \big/\, \text{\ion{C}{iii}}\,\lambda\, 5696$ line-strength ratio to drastically change as well, with similar trends visible for other diagnostic lines and line strength ratios.

\subsection{Continuum}\label{subsec:continuum}

\begin{figure}
    \centering
    \includegraphics[width=\hsize]{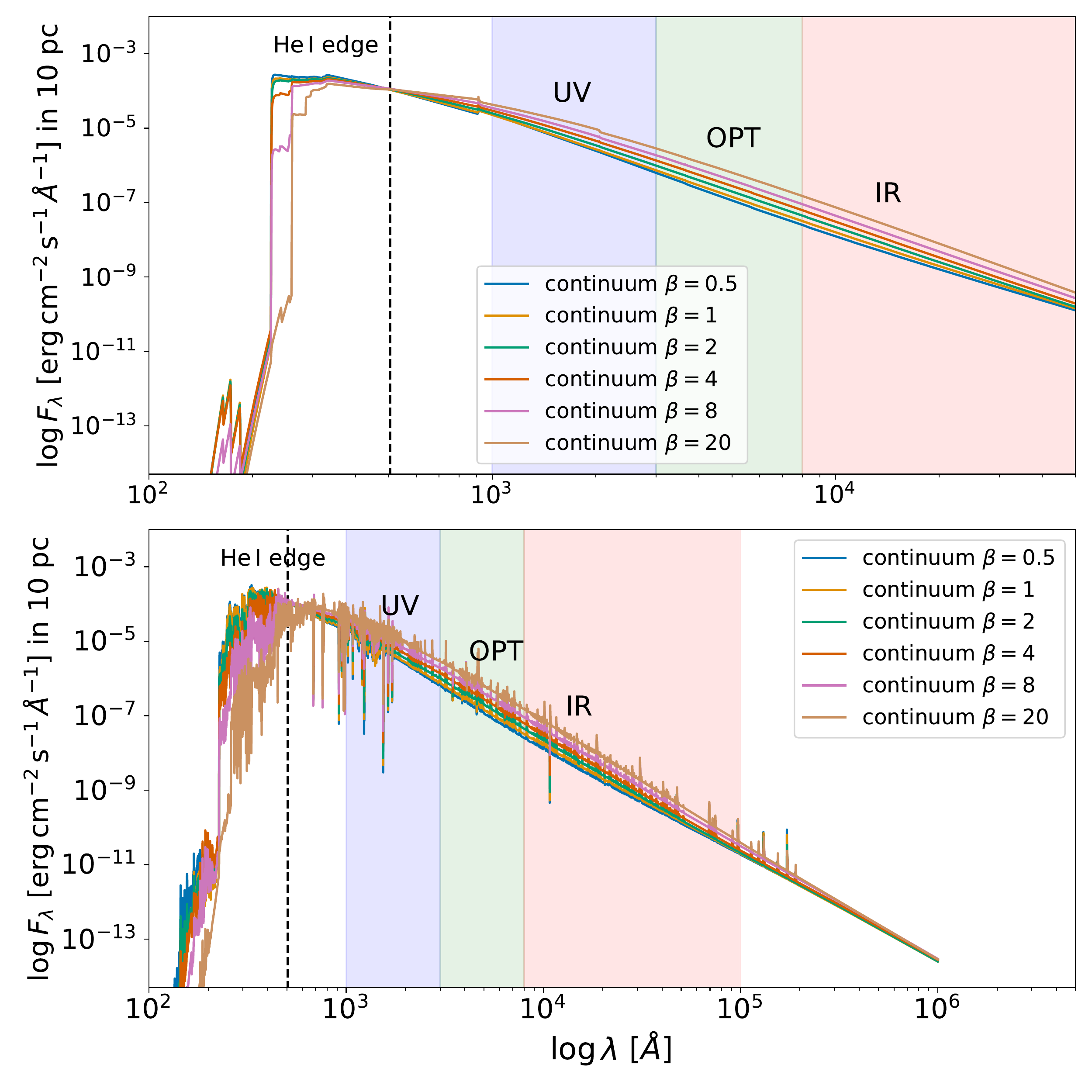}
    \caption{Continua of the B2 WN-star model for different $\beta$ velocity laws.
    The UV, optical and IR regimes are also highlighted.}
    \label{fig:sed_comp}
\end{figure}

In addition to the line spectrum, also the continuum itself is affected by changes of the velocity law. 
As an example, we take the B2 WN models (similarly to Fig.\,\ref{fig:wne_linewidths_comp}) and compare the continua for different $\beta$, shown in the upper panel of Fig.\,\ref{fig:sed_comp}.
The figure shows the continuum (without the lines) at a distance of $10\,\mathrm{pc}$ and is similar in shape to how it would be observed from Earth.
For all models in Fig.\,\ref{fig:sed_comp}, the total (bolometric) luminosity is the same. 
The initial impression seems to be that the continuum shifts to cooler temperatures with increasing $\beta$, an effect similar to the results above for the line strength ratios. 
Interestingly, the continuum of all models aligns at the \ion{He}{i} ionization edge ($504.3\,$\AA) in the extreme UV (EUV).
When looking towards longer wavelengths in the UV, optical, and infrared (IR) regimes, we further see that the flux is higher for higher values of $\beta$. 
In the optical regime, the shift is mostly wavelength-independent, corresponding to a simple shift in log-log space or a multiplication factor in linear units (see Sect.\,\ref{subsec:spectral_types} for further discussion). 
These continuum differences start to diminish for longer wavelengths, eventually converging around 100 $\mu$m.
When considering the full flux-calibrated spectrum (bottom panel of Fig.\,\ref{fig:sed_comp}), the differences are less obvious in the UV, but the overall trends remain. 
In the EUV, the differences become even more pronounced compared to the raw continuum case. 
All of our exemplary models, independent of $\beta$, are optically thick below the \ion{He}{ii} ionization edge, meaning that essentially no photons from this region escape from the stellar winds. 
However, the \ion{He}{i} ionizing flux is notably affected by the choice of the velocity field. 
For $\beta = 0.5$, the number of \ion{He}{i} ionizing photons is about an order of magnitude higher than in the $\beta20$ case. 
For the \ion{H}{i} ionizing flux ($\lambda < 912\,$\AA), the trend is more moderate with a decrease of  only $\approx$20\%. 
The trend is reversed for the $H_2$-dissociating Lyman-Werner photons, which are emitted between $912\,$\AA\ and $1110$\,\AA. 
For them, we obtain an increase of about a factor of two between $\beta = 0.5$ and $20$. 
However, the total photon budget below $1110$\,\AA\ remains approximately constant due to the different trends almost cancelling each other in our considered model sequence.

\subsection{Spectral classification}\label{subsec:class}

\begin{figure}
    \centering
    \includegraphics[width=\hsize]{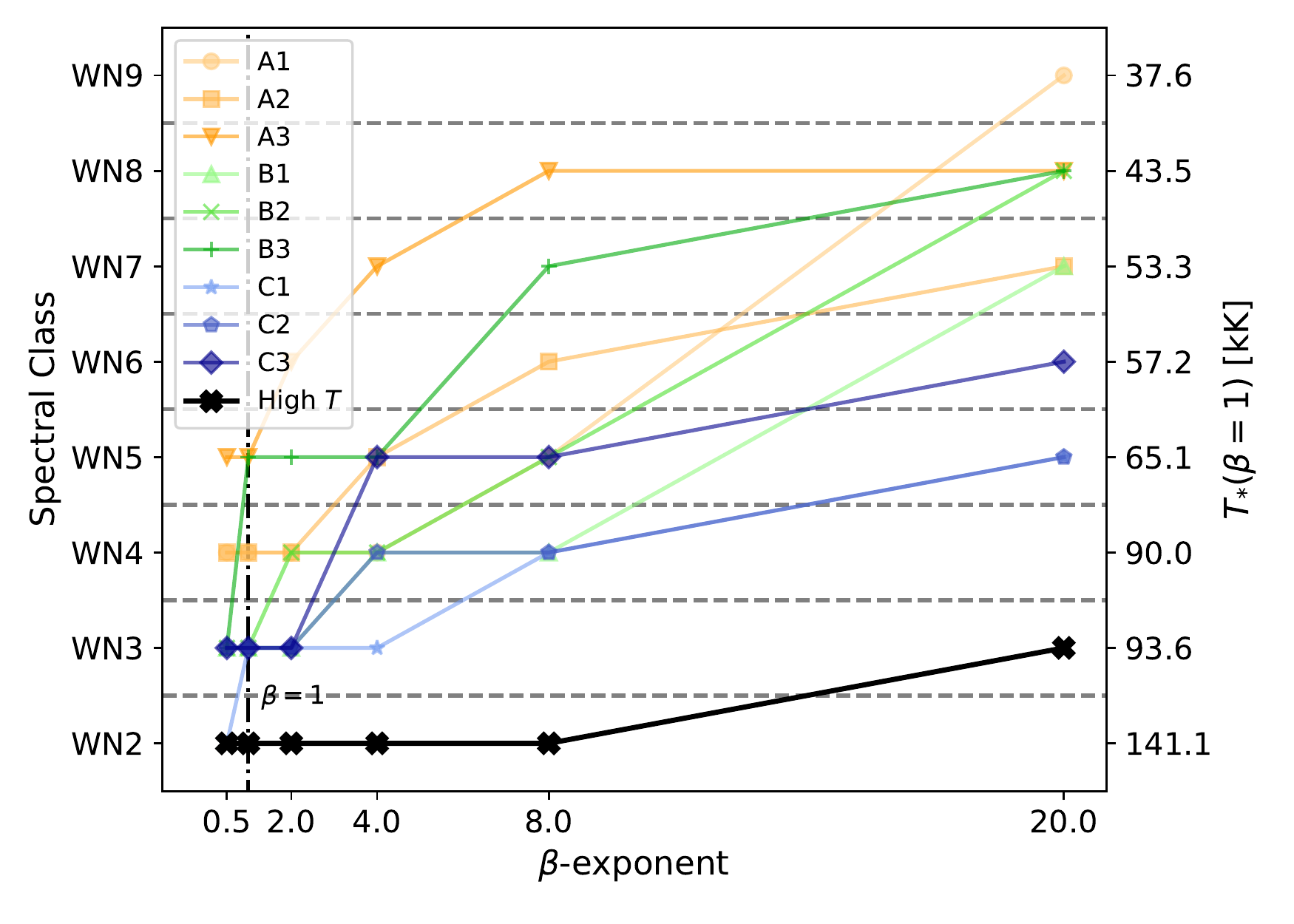}
    \caption{Spectral (sub)classification for the WN models (see Table \ref{tab:gridmodel_params} for parameters) as a function of $\beta$. 
    Note the trend to later-type spectral classes for increasing $\beta$-values, even for models with the same stellar parameters (e.g.\ the blue curve). 
    Additionally, a high $T_\ast$ model (see parameters in text) with temperatures comparable to the He-ZAMS in Fig.\,\ref{fig:comp} is classified for different $\beta$-values, to a much less large impact.}
    \label{fig:luka_classif_all}
\end{figure}

The spectral sub-classification for WR stars uses diagnostic line strengths and line strength ratios. 
With the strong influence on these lines and ratios by the $\beta$-parameter, the spectral subclasses are affected as well. 
In Fig.\,\ref{fig:luka_classif_all}, we show the classification -- here based on \citealt{smith1996three} -- of nine WN-star models with different stellar parameters (see Table \ref{tab:gridmodel_params}) as a function of the $\beta$-parameter.
For observed WR stars, the early and late-type spectral subclasses are commonly interpreted as an indicator for the stellar temperature $T_\ast$ at the hydrostatic layers: early-type classes are associated with hotter WR stars and later-type classes with cooler WR stars. 
In our $\beta$-study, we now find for all nine WN-star model sequences, the spectra are classified as a later subtype when increasing the $\beta$-parameter compared to their classification when $\beta=1$ is used.

As an example, the $\beta1$ B2 model in Fig.\,\ref{fig:luka_classif_all} is classified as an early-type WN3 star, while the $\beta20$-model is classified as a late-type WN8 star despite the stellar parameters being identical, save for $\beta$.
Moreover, the $\beta20$ C1 and $\beta1$ A3 models are both classified as intermediate-type WN5 stars, despite a temperature difference of $\Delta T_\ast \approx 40$\,kK between the two models.
Given that large model grids are so far only available for $\beta1$ laws, one could associate average temperatures with accompanying spectral subclass. 
This has been done in the right-hand side axis in Fig.\,\ref{fig:luka_classif_all}, where we average over the results from \citealt{hamann2006galactic}, employing $\beta1$ models.
An overview of these average temperatures is also shown in Table \ref{tab:temps_specclass}.
Our compiled $T_\ast(\beta=1)$ values essentially reflect which temperatures would be inferred for a specific spectral subclass when employing only $\beta1$-models.
Assuming that $\beta$ values are also distributed over a range of values in nature, this implies that there is no coherent relation between spectral subtypes and hydrostatic temperatures of WR stars.

In addition to the nine model sequences representing the bulk of observed WN stars, we further calculated a series of additional (denoted as high $T_\ast$ in Fig.\,\ref{fig:luka_classif_all}) models with different $\beta$-laws representing a WN star close to the He-ZAMS (cf.\ Fig.\,\ref{fig:comp}). 
Motivated by the steady-state model from \citet{poniatowski2021dynamically}, the stellar parameters of this model sequence are $T_\ast = 130$ kK, $\log(\dot{M}_\mathrm{t}\,[M_\odot/\mathrm{year}]) = -4.434$, $R_\mathrm{t} = 2.0893\, R_\odot$, $\log(L_\ast\,[L_\odot]) = 5.416$ and $\log(\dot{M}_\ast\,[M_\odot/\mathrm{year}]) = -4.831$.
Interestingly, the spectral classification of the high $T_\ast$ series is less affected by the different $\beta$-values in the stellar wind than the other, lower-$T_\ast$ models, as evident from Fig.\,\ref{fig:luka_classif_all}.
Here, the different models are classified as WN2 for all $\beta$-values with the exception $\beta = 20$. 
However, the value of $\dot{M}_\text{t}$ in the model motivated by \citet{poniatowski2021dynamically} is lower than in all of our other models, so we cannot rule out that the effect might be stronger for higher mass-loss rates (i.e.\ even more dense winds).

\begin{figure}
    \centering
    \includegraphics[width=\hsize]{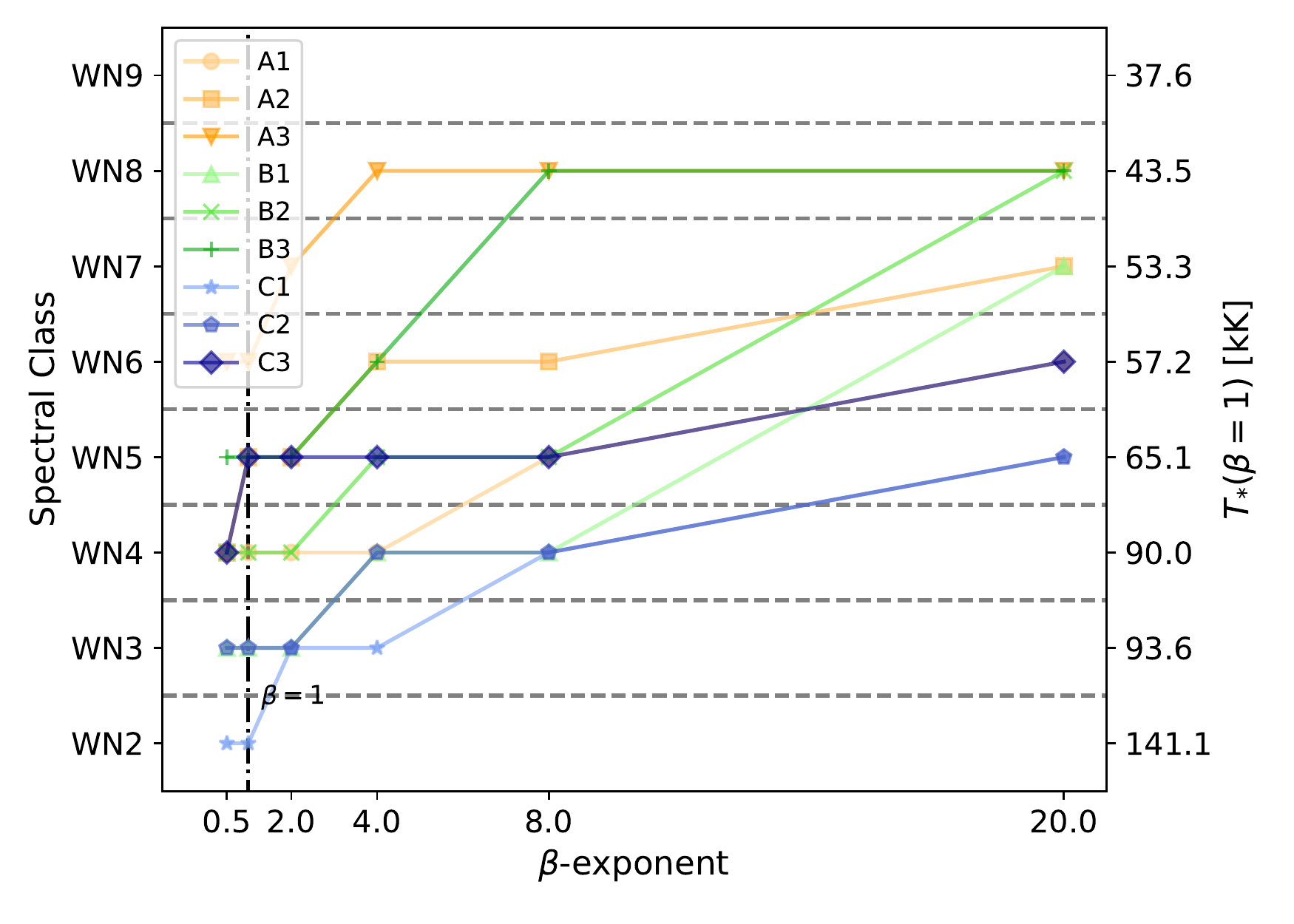}
    \caption{Similar to Fig.\,\ref{fig:luka_classif_all} but now for the WNh models (see Table \ref{tab:gridmodel_params}). 
    Here too there is a general trend to later-type spectral classes for increasing $\beta$-values.}
    \label{fig:wnlh_classif_all}
\end{figure}

The same analyses have been performed on the WNh-type models, of which the results are shown in Fig.\,\ref{fig:wnlh_classif_all}.
While there are differences in the individual classifications, we see the same effect of increasing $\beta$ as in the WN sequences: models with higher $\beta$-values in their stellar winds tend to be classified as later-type stars. 
This thus leads to the same ambiguity between the apparent $T_\ast$ and the $\beta$-value.

\begin{figure}
    \centering
    \includegraphics[width=\hsize]{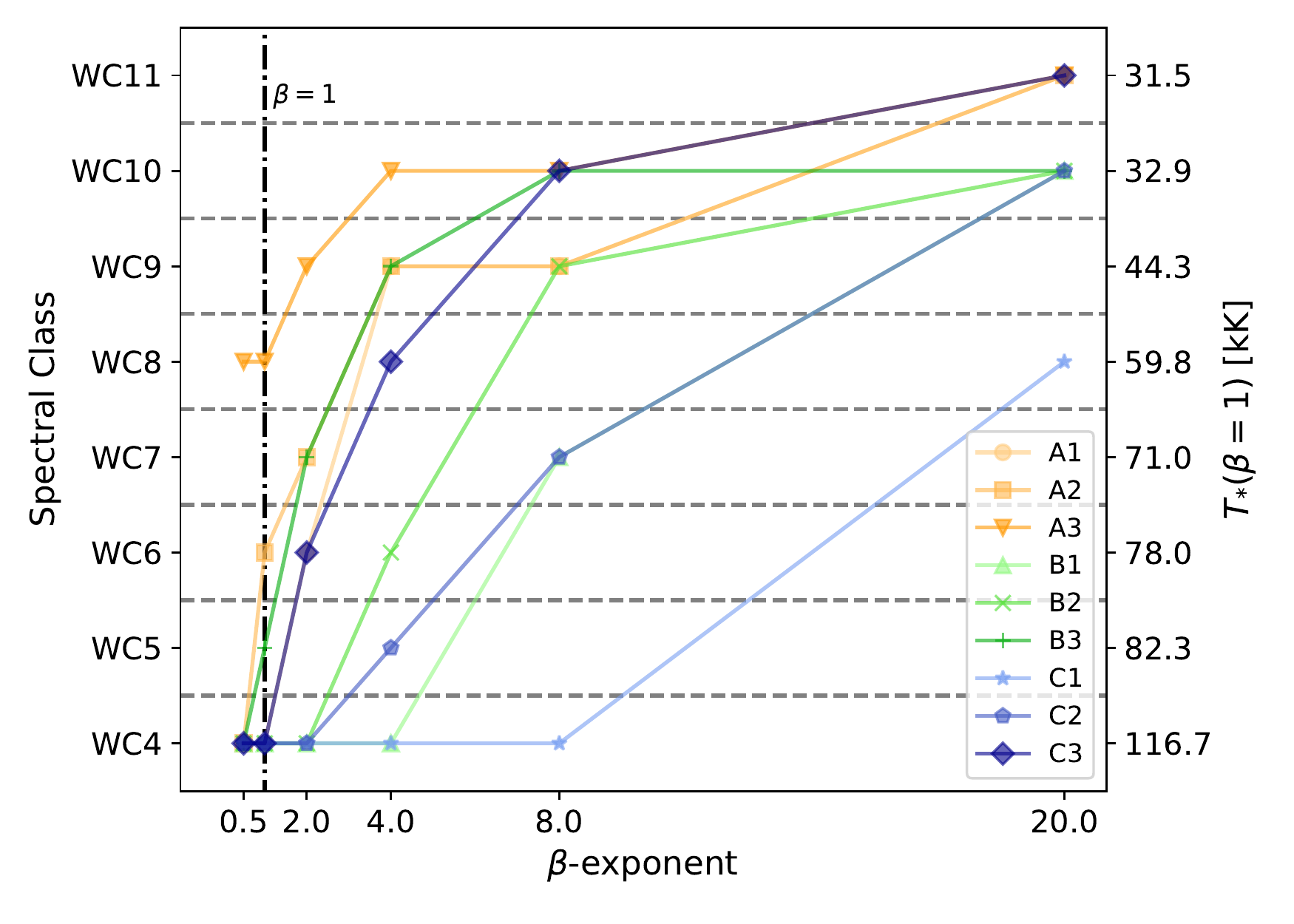}
    \caption{Similar to Figs.\,\ref{fig:luka_classif_all} and \ref{fig:wnlh_classif_all} but now for the WC models (see Table \ref{tab:gridmodel_params}. 
    A similar trend to later spectral types for increasing $\beta$ can be seen.}
    \label{fig:wc_classif_all}
\end{figure}

Unsurprisingly, the classification of the WC-star models is affected as well. 
Applying a similar analysis as for the WN-type models in this work, but now employing the classification criteria from \citet{crowther1998quantitative}, the resulting subtypes are shown in Fig.\,\ref{fig:wc_classif_all}. 
As evident from the figure, the model spectra tend to be again classified as later-type stars for higher $\beta$-values.
The right-hand axis in Fig.\,\ref{fig:wc_classif_all} shows the stellar temperatures per subclass (compiled from \citealt{sander2012galactic} and \citealt{leuenhagen1996spectral}, see also Table \ref{tab:temps_specclass}).
This again yields the same ambiguity in stellar temperature and the $\beta$-parameter as with the WN-star models.

\begin{table*}
    \caption{Average values of derived temperatures of observed WR stars using $\beta1$-models. 
    The WN-star values are compiled from \citet{hamann2006galactic}, \citet{Oskinova+2013}, and \citet{Sander+2014}, while the WC-star values originate from \citet{LeuenhagenHamann1998}, \citet{deMarcoCrowther1998} and \citet{sander2012galactic}.
    }
    \centering
    \vspace{2mm}
    \def\arraystretch{1.3} % increase column height
    \begin{tabular}{|cc|cccccccccc|}
        \hline
        \multicolumn{2}{l}{Spectral class:} & 2 & 3 & 4 & 5 & 6 & 7 & 8 & 9 & 10 & 11\\
        \hline
        WN & $T_\ast$ [kK]$^a$ & $141.1$ & $93.57$ & $89.98$ & $65.09$ & $57.24$ & $53.28$ & $43.52$ & $37.65$ & $25.75$ & $22.39$ \\
        & $T_{2/3}$ [kK]$^a$ & $104.7$ & $83.2$ & $54.1$ & $47.7$ & $44.8$ & $42.6$ & $37.4$ & $36.7$ & $20.4$ & $19.1$ \\
        WC & $T_\ast$ [kK] & & & $116.7$ & $82.3$ & $78.0$ & $71.9$ & $59.8$ & $44.3$ & $32.85$ & $31.5$\\
        & $T_{2/3}$ [kK] & & & $73.0$ & $63.1$ & $62.0$ & $61.2$ & $52.2$ & $39.0$ & $30.45$ & $29.5$ \\
        \hline
        \multicolumn{11}{l}{$^a$The same temperatures apply for the WNh-star models.}
        % \multicolumn{11}{l}{$^b$No $T_{2/3}$ values known for WC10 and WC11.}
    \end{tabular}
    \label{tab:temps_specclass}
\end{table*}

\subsection{Line profile changes}\label{subsec:changes}

\begin{figure}
    \centering
    \includegraphics[width=\hsize]{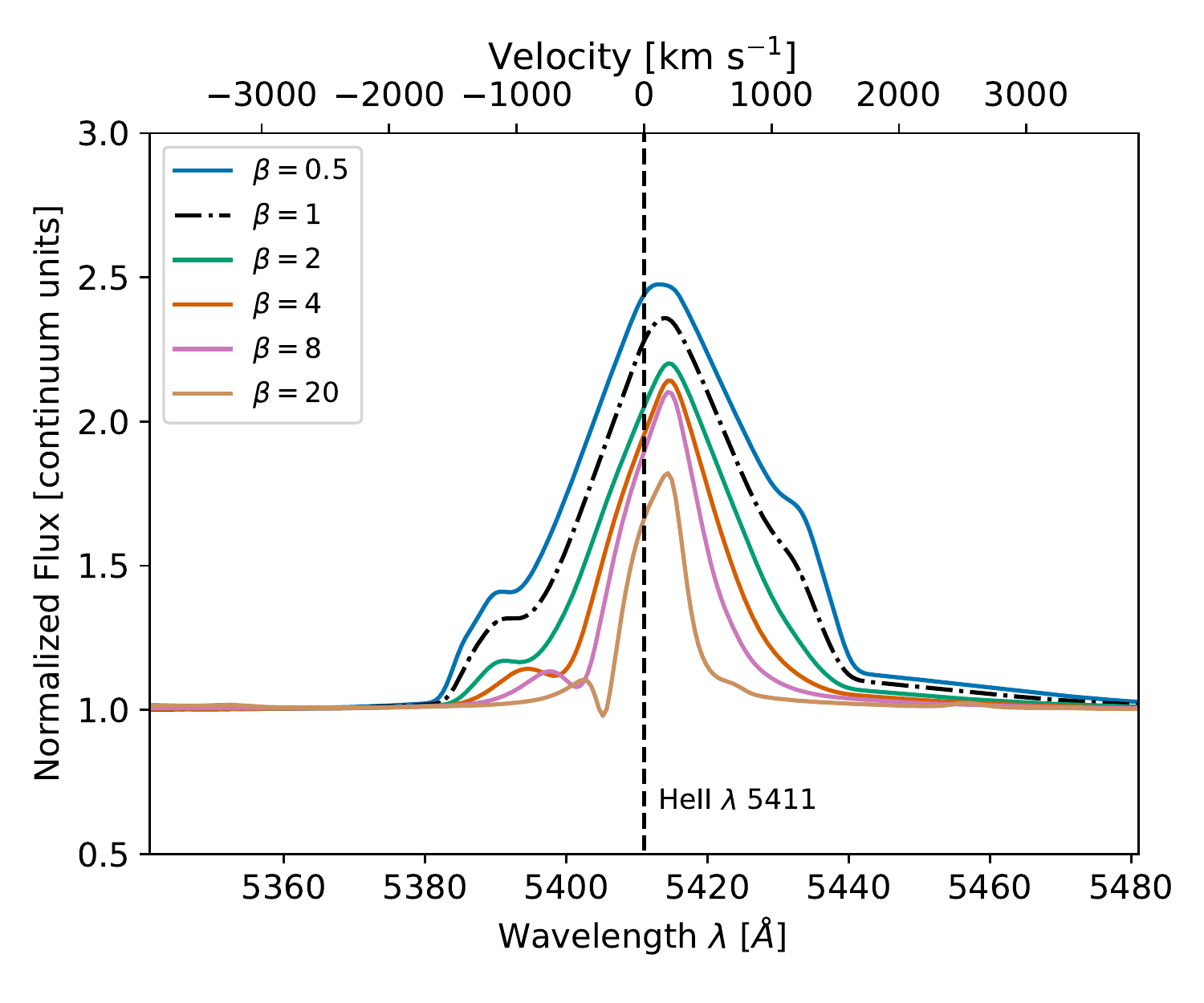}
    \caption{The optical \ion{He}{ii} $\lambda$ 5411 diagnostic line is shown for the WN B2 models with different $\beta$-values.
    The width of the line is strongly affected by the $\beta$-parameter used in the velocity law: increasing $\beta$-values lead to narrower lines.
    This effect is visible for other optical spectral lines as well.}
    \label{fig:opt_line}
\end{figure}

Beside affecting peak heights, the use of different $\beta$ values for $\varv(r)$ also leads to different line widths. 
An example is shown in Fig.\,\ref{fig:opt_line}, where the \ion{He}{ii} $\lambda$ 5411 spectral line is shown for the WN B2 models.
The velocities associated with the Doppler-shift and accompanying line broadening from the center of the line are shown on top.
The figure shows a strong decrease in both line-strength (peak-to-continuum value) and line width, despite the $\dot{M}$ and $\varv_\infty$-values being the same for the different spectra.

Spectral classification of WN stars sometimes also incorporates line widths, discerning between strong and weak-lined spectra. 
As a measure for the line widths, \citet{smith1996three} uses the full width at half maximum (FWHM) of the \ion{He}{ii} $\lambda\, 4686$ line, where thresholds of $\mathrm{FWHM} > 30$ \AA\, and $\mathrm{FWHM} < 30$ \AA\, are used for broad-line and weak-lined spectra respectively. 
While the measure as such is purely morphological, our study highlights that any immediate conclusions about the overall wind strength or mass loss of the objects based only on emission-line widths are ambiguous. 
Without additional knowledge on their formation regions, terminal wind velocities measured from the width of emission lines need to be considered as lower limits.
This is due to the strong influence of the velocity field on the line-widths. In case of $\beta$-type laws, weaker emission lines can even be accompanied by relatively high mass-loss rates if $\beta \gg 1$.

\subsection{UV spectra}\label{subsec:uv}

\begin{figure}
    \centering
    \includegraphics[width=\hsize]{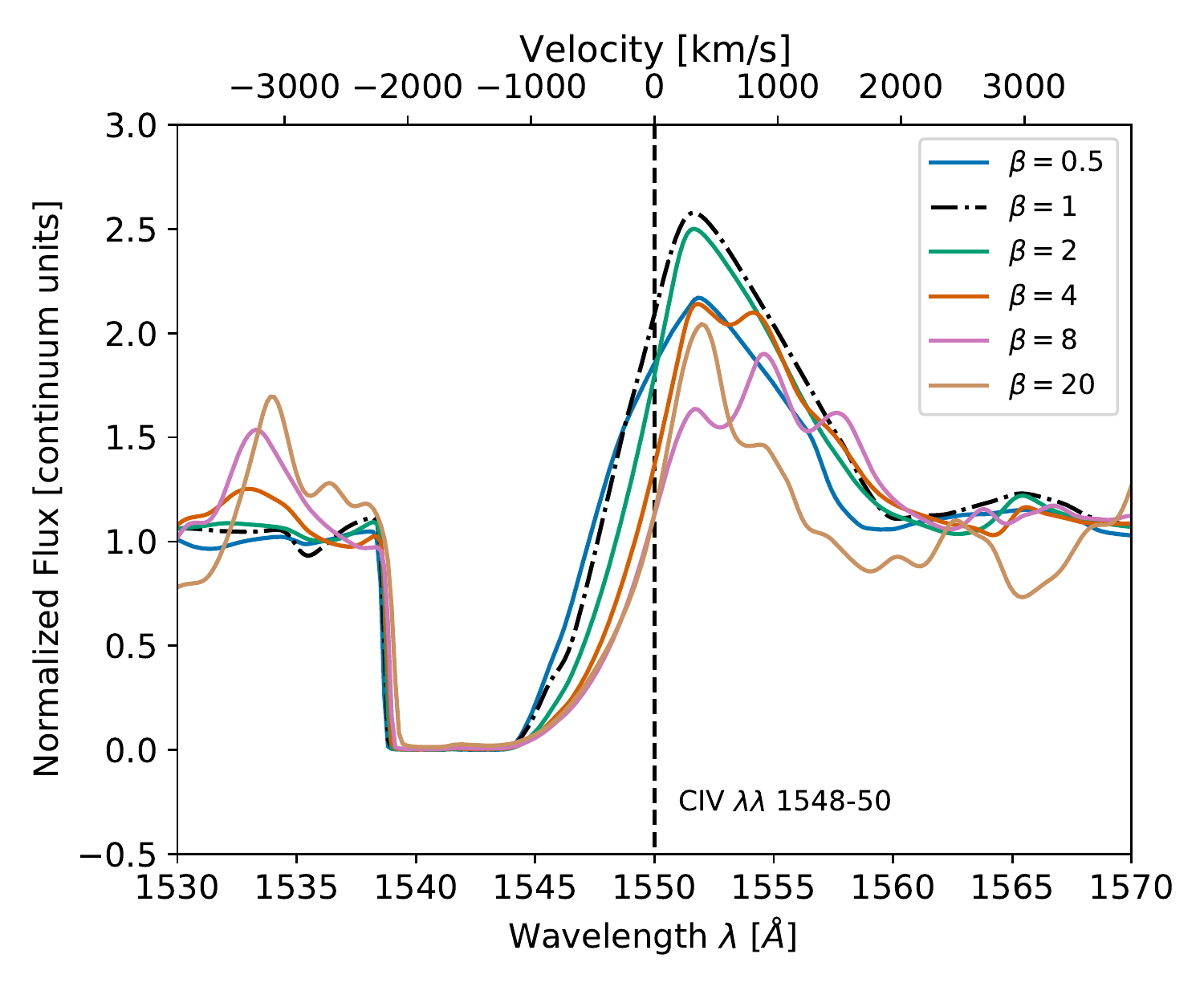}
    \caption{The UV \ion{C}{iv} $\lambda\lambda$ 1548-50 P-Cygni profile is shown for the WN B2 model with different $\beta$-values. 
    Although the emission part of the P-Cygni line varies strongly with the value for $\beta$, the blue-most part of the absorption trough remains largely unaffected.}
    \label{fig:uv_line}
\end{figure}

The main lines used for classification of WR stars are located in the optical spectrum. 
At Galactic metallicities, the optical spectrum of WR stars is usually dominated by emission lines. 
This is not necessarily the case for the UV spectrum. 
Given the previously discussed ambiguities for deriving the wind parameters only from (optical) emission lines, we now take a look at the effect of different $\beta$-law assumptions on prominent UV lines. 
As a case study, we take the \ion{C}{iv} $\lambda\lambda$ 1548-50 doublet -- a P-Cygni profile -- shown in Fig.\,\ref{fig:uv_line} for the WN B2 models, applying different $\beta$-values.
The P-Cygni profile in the figure shows a varying emission part, similar to the optical emission lines, while the blue-most wavelength of the P-Cygni absorption part remains unaltered when changing the $\beta$-parameter. 
Similar results are obtained for other WN and WC star models.

\begin{figure*}
    \centering
    \includegraphics[width=\hsize]{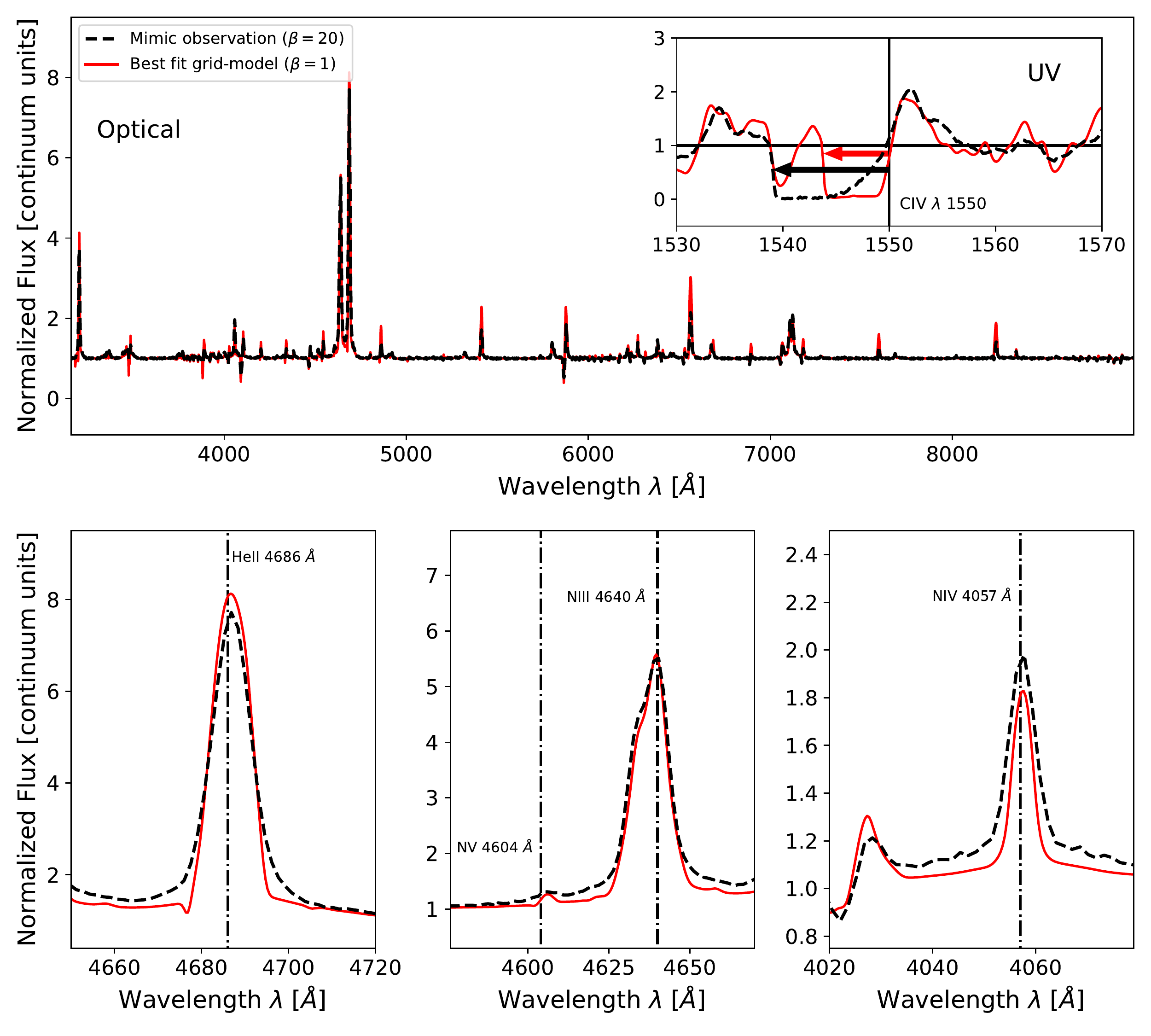}
    \caption{The main graph on top shows a comparison of optical model spectra for a simulated observation (with $\beta=20$, $R=3600$, $\mathrm{SNR}=100$) and a best-fit grid-model (with $\beta=1$), in black and red respectively (see text for parameters).
    Optically, both model spectra resemble each other closely, while strong discrepancies arise in UV P\,Cygni lines, such as the \ion{C}{iv} $\lambda\lambda$ 1548-50 doublet shown in the inset.
    The bottom three plots show an enlargement of some of the diagnostic lines (see also Table \ref{tab:diagnostic_lines}).}
    \label{fig:mimick}
\end{figure*}

To illustrate the ambiguity caused by varying the $\beta$-parameter and the potential of UV data in removing it, we mimic an observation in the visual domain by degrading the resolution of an emergent PoWR spectrum. 
Additionally, Gaussian noise is added to simulate a finite signal-to-noise (SNR)-ratio.
The simulated observation used here is a WN B2 (see Table \ref{tab:gridmodel_params}) $\beta20$-model with $\varv_\infty = 1600$ km s$^{-1}$, degraded to a resolution of $R = 3600$ in similarity to observational data used in \citet[e.g.][]{hamann2006galactic} -- originally from \citet{1995A&AS..113..459H} -- and with a simulated $\mathrm{SNR} = 100$.
We then proceed by comparing PoWR $\beta1$ models to the mock observation considering only the optical regime and compare the \ion{C}{iv} $\lambda\lambda$ 1548-50 doublets of both the mock observation and the best fitting model.
The best-fit $\beta1$ model is a WN model with $T_\ast = 44.7$ kK, $\dot{M}_\mathrm{t} = -4.06$ and $\varv_\infty = 600$ km s$^{-1}$. 
Both are shown in Fig.\,\ref{fig:mimick}, along with a zoom-in on diagnostic lines and the \ion{C}{iv} $\lambda\lambda$ 1548-50 profile.

While we do not aim at a fine-tuned analysis here -- a few optical lines do not agree perfectly -- the optical spectra in the figure resemble each other reasonably well in the context of a grid-based approach.
Such an approach would be used when studying a larger sample or using a grid model to account for the WR imprint in an unresolved population. 
On the contrary, the P-Cygni profile shows a large discrepancy in the blue-most wavelength of the absorption trough. 
Moreover, there are significant differences in other stellar parameters between both models: a difference in $T_\ast$ of 26.1 kK and a $\dot{M}_\mathrm{t}$ difference of 0.3 dex (the true $\dot{M}$ difference is 0.12 dex). 
Thus, beside underestimating $\varv_\infty$ by $1000$ km s$^{-1}$, the $\beta1$-model further severely underestimates $T_\ast$ and underestimates $\log \dot{M}$ of the simulated observation. 
We also note that the difference in $T_{2/3}$ between these models is 6.8 kK, noticeably smaller than the $T_\ast$ difference and more in line with the similarity of the line spectra.
Also the continuum differences are fairly small. 
While there is some discrepancy in the optical and infrared regimes, in a similar way as in Fig. \ref{fig:sed_comp}, the effect is much smaller and amounts only to an apparent luminosity difference of $\sim 0.004$ dex in the optical regime (while the real model luminosities are the same).
Lastly we want to highlight that the differences in the \ion{C}{iv} $\lambda\lambda$ 1548-50 absorption troughs are not caused by the different velocity laws, in contrast to Fig. \ref{fig:uv_line}, but due to different $\varv_\infty$ in the input of the models.

\subsubsection*{Implications for stellar feedback}

Underestimations of the terminal velocity can also have implications for the derived stellar feedback. 
Regarding the ionizing feedback, the $\beta1$ model has $\sim 10$\% more hydrogen ionizing flux than the mock observation.
When considering the mechanical luminosities
\begin{equation}
  L_\mathrm{mech} = \frac{1}{2}\dot{M}\varv_\infty^2,
\end{equation}	
the consequences of underestimating $\varv_\infty$, e.g.\ due to considering only $\beta 1$-velocity law, can be significant as well. 
In our example, the $\sim 35$\% difference in $\dot{M}$ and the 2.7 times higher $\varv_\infty$ both need to be taken into account, resulting in an underestimation of $L_\mathrm{mech}$ in our mock observation in Fig.\,\ref{fig:mimick} by nearly an order of magnitude if the values from the $\beta1$ ``fit'' are adopted.

\subsection{Comparison to observations}\label{subsec:obs}

To get a first look on whether our findings from the model calculations are realized in nature, we take three objects with available optical and UV observations as case studies: WR128, SMC\,AB2 and M31\,WR096 (LGGS\,J004412.44+412941.7).
These targets were chosen due to the availability of UV spectra as well as their difference in metallicities and galactic backgrounds (Milky Way, SMC and M31, respectively).
Their derived properties are shown in Table \ref{tab:obsdat_stars}, where the data of the analysis of WR128 from \citet{hamann2019galactic} and of SMC\,AB2 from \citet{Hainich+2015} are also given for comparison. 
In both cases, a $\beta1$-velocity law was used to determine the stellar parameters in the original work.
For SMC AB2, we use the stellar parameters found by \citet{Hainich+2015} as the best $\beta1$-representation. 
For WR128 the terminal velocity was not derived by custom modelling in \citet{hamann2006galactic} and we thus calculated a dedicated $\beta1$-model with an adjusted $\varv_\infty$ for WR128. 
The same was done for M31 WR096, where no previous analysis has been performed. 
Without aiming for fine-tuned spectral fits for each of the targets, which would require a separate dedicated study, we calculate models with $\beta \neq 1$ and compare them to the best $\beta1$-solution.
The detailed results for three targets are discussed in Appendix \ref{sec:appendix}. 
We give a short overview here as well.

For all three targets -- WR128, SMC\,AB2 and M31\,WR096 -- $\beta4$ models seemed to give the best agreement with the observed spectra. 
Higher $\beta$ velocity laws were increasingly difficult to reconcile with the observed spectra.
Given the UV lines, the spectrum of WR128 is reproduced better with a $\beta4$ velocity law, while the optical regime showed similar agreement between the $\beta1$, $\beta4$ models and the observation. 
The $\beta4$ model for SMC AB2 shows a comparable quality to the $\beta1$ model. 
However, an unfeasibly large clumping factor of $D=130$ was needed to attain this. 
Additionally, the observed UV spectrum is also recreated well with a $\beta1$ velocity law, suggesting the original $\beta1$ model should be used in this case.
In a similar way to WR128, also for M31 WR096 the $\beta4$ model can be favoured based on the slightly better quality of the UV spectrum.

With these three targets we investigated whether models with $\beta \neq 1$ could generally reproduce the spectrum in a similar quality as a ``traditional'' $\beta1$ model. 
We found that for two of the three targets one could justify a $\beta4$ velocity law, while one target (SMC\,AB2) was best reproduced with the $\beta1$-model. 
As models with even higher $\beta$ values yielded less good results, we thus conclude that stellar models with $\beta>20$ will only rarely, if at all, provide a suitable representation of existing WR winds. 
This, fortunately, limits the impact of the aforementioned degeneracy with the $\beta$ law. Nonetheless, a range of $\beta$ values around $1$ should be explored when analysing WR stars with prescribed $\beta$-law. 
To determine the ``best'' $\beta$-value, optical data is only of limited help. Instead, the saturated P-Cygni profiles in UV, which are still available for most WR stars, are important to not underestimate the value $\varv_\infty$.

Considering the derived parameters in Table \ref{tab:obsdat_stars}, we see that for stars analysed with different $\beta$-models, the most striking differences are seen between the inferred $T_\ast$ and $\varv_\infty$ values, with noticeable but less severe discrepancies between the $\dot{M}$ values. 
On the other hand, while the $T_{2/3}$ values are hardly affected by the differences in wind parameters, there can be significant changes in the luminosity, when other parameters are adjusted simultaneously with $\beta$. 
This is reflected most clearly in our analysis of SMC AB2 (see Sect. \ref{sec:appendix}).
As the other two studied targets are hardly affected, we conclude that different choices of $\beta$, within the acceptable values for observations, only have a mild effect on the resulting spectral energy distribution (SED).
The similarity of the $T_{2/3}$ temperatures could be attributed to the motivation of getting the same spectral appearance.
This would require that the shape of the velocity field hardly affects $T_{2/3}$, which seems to be valid for $\beta$-type velocity laws, but will not hold in general as we will show later in Sect.\,\ref{sec:hydrocmp}.

\begin{table*}
    \caption{Inferred parameters from previous spectral fitting of WR128 \citep{hamann2019galactic} and SMC\,AB2 \citep{Hainich+2015} (original fits) and of models fitted in this study.
    M31\,WR096 has no inferred stellar parameters from previous studies.
    Noticeable here are the discrepancies between $T_\ast$ and $\varv_\infty$ from models with different $\beta$-values.}
    \label{tab:obsdat_stars}
    \centering
    \def\arraystretch{1.3}
    \begin{tabular}{|c|cccccccccccc|}
        \hline
        Target & $T_\ast$ & $T_{2/3}$ & $\log R_\text{t}$  & $\varv_\infty$ & $X_\mathrm{H}$ & $E_{b-\varv}$ & Law$^a$ & $R_\ast$ & $R_{2/3}$ & $\log \dot{M}^b$ & $\log L$ & $\beta$\\
         & [kK] & [kK] & $[R_\odot]$ & \!\![km\,s$^{-1}$]\!\! & [\%] & [mag] & $R_V$ & $[R_\odot]$ & $[R_\odot]$ & $[\dot{M}\,\mathrm{yr}^{-1}]$ & $[L_\odot]$ & \\
        \hline
        WR128 \citep{hamann2019galactic} & $70.8$ & $70.0$ & $1.1$ & $2050$ & $16$ & $0.32$ & C $3.6$ & $2.69$ & $2.73$ & $-5.4$ & $5.22$ & $1$\\
        WR128 (this study, $\beta=1$) & $70.8$ & $69.2$ & $1.1$ & $1700$ & $50$ & $0.38$ & C $3.6$ & $3.04$ & $3.18$ & $-5.32$ & $5.42$ & $1$\\
        WR128 ($\beta\neq 1$) & $89.2$ & $65.9$ & $0.9$ & $2500$ & $50$ & $0.30$ & C $3.6$ & $1.86$ & $3.41$ & $-5.24$ & $5.25$ & $4$\\
        SMC\,AB2 \citep{Hainich+2015} & $47.3$ & $47.1$ & $1.63$ & $900$ & $55$ & $0.10$ & G $2.74$ & $9.10$ & $9.17$ & $-5.75$ & $5.57$ & $1$\\
        SMC AB2 ($\beta\neq 1$) & $56.2$ & $52.9$ & $1.7$ & $1100$ & $55$ & $0.16$ & G $2.74$ & $6.65$ & $7.26$ & $-6.45$ & $5.97$ & $4$\\
        M31 WR096 ($\beta=1$) & $63.1$ & $55.8$ & $0.7$ & $2500$ & $0$ & $0.15$ & F $3.1$ & $3.90$ & $4.92$ & $-4.66$ & $5.6$ & 1\\
        M31 WR096 ($\beta\neq 1$) & $89.1$ & $52.2$ & $0.4$ & $3000$ & $0$ & $0.12$ & F $3.1$ & $1.98$ & $5.56$ & $-4.46$ & $5.5$ & $4$\\
        \hline
        % \multicolumn{12}{l}{$^a$ The temperature at continuum Rosseland mean opacity $\tau_\mathrm{Ross} = 2/3$.}\\
        \multicolumn{13}{l}{$^a$Reddening law: C stands for \citet{1989ApJ...345..245C}, G for \citet{2003ApJ...594..279G} and F for \citet{fitzpatrick1999correcting}, along with the $R_V$-values.}\\
        \multicolumn{13}{l}{$^b$The WR128 models and the $\beta1$ SMC AB2 model use clumping factor $D=4$, the $\beta4$ SMC AB2 model has $D=130$,}\\
        \multicolumn{13}{l}{the M31\,WR096 models use $D=10$.}
    \end{tabular}
\end{table*}

\section{Discussion: The $\beta$-approach for WR stars}\label{sec:discussion}

The differences in the emergent spectra seen in Sect. \ref{sec:results} can physically be attributed to the changes in the different $\beta$-velocity fields, shown in Fig.\,\ref{fig:betalaws}, and to their coupling with densities and temperatures (see Sect. \ref{subsec:spectral_types} and further) in the stellar wind.

\subsection{Determination of terminal velocities from line widths}\label{subsec:widths}

\begin{figure}
    \centering
    \includegraphics[width=0.9\hsize]{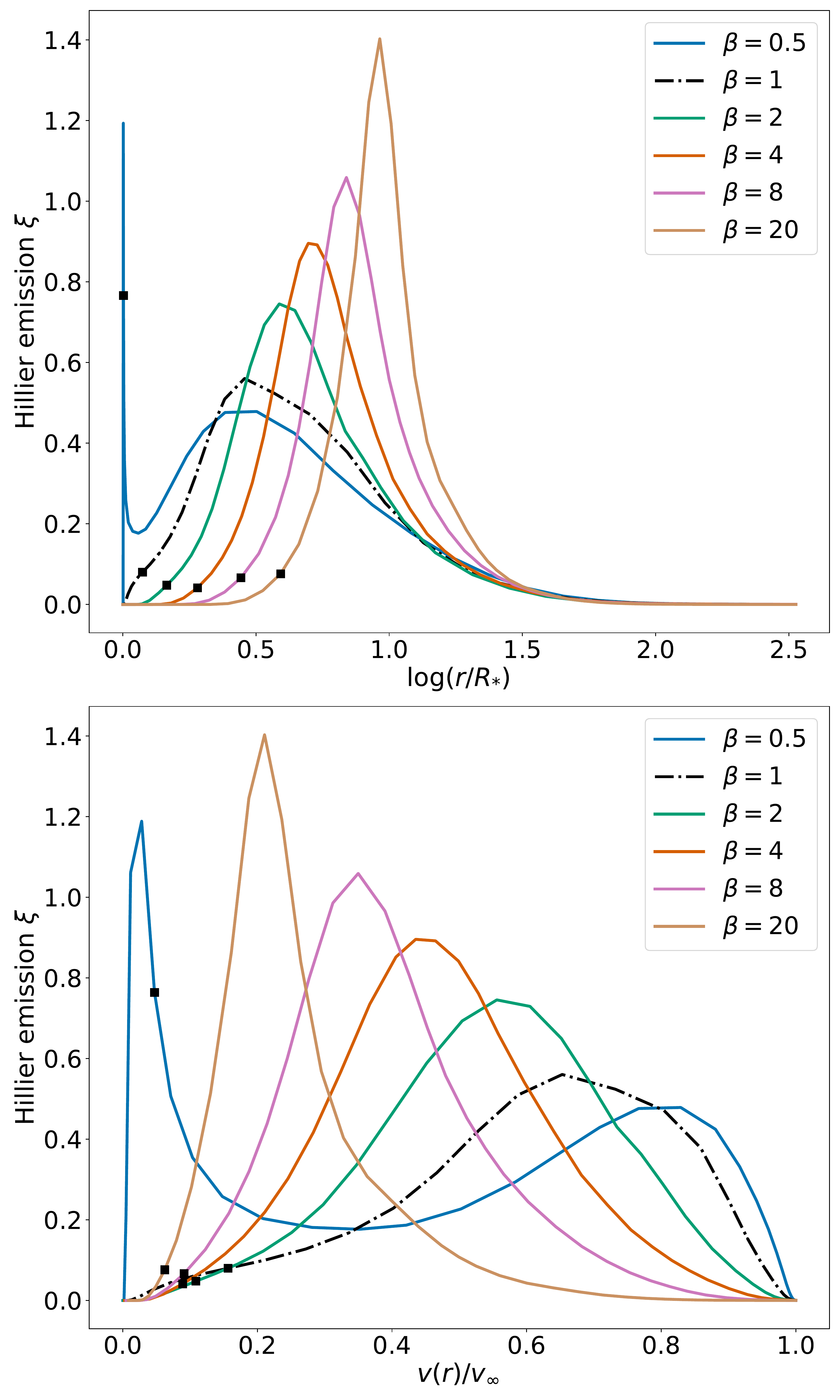}
    \caption{Emergent line emission $\xi$ \citep[from][]{1987ApJS...63..965H} as a function of the stellar radius (top) and of the wind velocity (bottom) of the B2 WN-star model. 
    In these plots, we show the computations for the \ion{He}{ii} $\lambda$ 5411 line from Fig.\,\ref{fig:opt_line} for models with different $\beta$-values.
    The first peak at low radii and low velocities for the $\beta0.5$-model (light blue line) has a negligible contribution to the actual emission line.
    The black squares denote the velocities and radii where $\tau_\mathrm{Ross}=2/3$.}
    \label{fig:hillier_emission}
\end{figure}

The change in diagnostic line widths, shown in e.g.\ Figs.\,\ref{fig:wne_linewidths_comp}, \ref{fig:wc_linewidths_comp}, and \ref{fig:opt_line}, has implications in determining the correct $\varv_\infty$.
Physically, the different velocity fields can cause the line formation regions to shift closer to or further away from the star.
In the case of the $\beta$-law, velocity descriptions with higher $\beta$-values have a shallower velocity increase.
Hence, higher velocities are reached at larger radii from the star.
This is demonstrated in Fig.\,\ref{fig:hillier_emission} for the example of the \ion{He}{ii} $\lambda$ 5411 line, depicted in Fig.\,\ref{fig:opt_line} where a measure of the line-formation strength $\xi$ \citep[from][]{1987ApJS...63..965H} is used.
While the value of $\xi$ does not reflect the intrinsic line strength, its distribution reflects where an emission line forms in the wind.
The plot in Fig.\,\ref{fig:hillier_emission} clearly shows that the maximum of $xi$ is shifting to larger radii but lower wind velocities when increasing the $\beta$-value.
When the line emission happens at lower velocities, the corresponding lower Doppler shifts lead to less broadened spectral lines, as also evident from Fig.\,\ref{fig:opt_line}.
Consequently, the derived $\varv_\infty$ could be ambiguous if only inferred from the optical spectrum.
For each of the $\beta$-models in Fig.\,\ref{fig:hillier_emission}, also the radii and velocities where $\tau_\mathrm{Ross}=2/3$ are indicated. 
We see the $R(\tau_\mathrm{Ross}=2/3) = R_{2/3}$ consistently moving outwards for increasing $\beta$, while $\varv(\tau_\mathrm{Ross}=2/3) = \varv_{2/3}$ peak at $\beta = 1$ with then an overall decreasing trend for higher $\beta$.
 
This problem can be resolved by using saturated P-Cygni lines in the UV-spectrum, most notably the \ion{C}{iv} $\lambda\lambda$ 1548-50 doublet. 
As the emission component of these lines form relatively close to the stellar surface, the absorption troughs are expected to be strong enough to overcome the emission in the P-Cygni profile, although this emission can still vary strongly with different velocity laws. 
For example, the P-Cygni profile in Fig.\,\ref{fig:uv_line} is saturated and thus gives a suitable estimate for $\varv_\infty$ \citep{1990ApJ...361..607P} independent of the velocity law used in modelling the stellar wind. 
Moreover, fixing $\varv_\infty$ via the UV P-Cygni absorption trough allows to exploit the $\beta$-dependence of the optical diagnostic lines to estimate the best $\beta$-value. 
We briefly tested this in our observational examples, although a broader range of model calculations would be required to give a quantitative estimate of the best $\beta$-value and its error margin.

For O-type stars, UV P-Cygni profiles have been used extensively to determine the terminal wind velocity as typically most distinctive wind signatures are found in the UV \citep[e.g.][]{1990ApJ...361..607P}. 
As WR-star optical spectra usually contain plenty of emission lines formed in the wind of the star, these have typically been used over UV P-Cygni profiles to constrain $\varv_\infty$.
The above results showing the optical line width ambiguity for the $\beta$-parameter now stress the importance of UV P-Cygni lines also for WR stars.

\subsection{Degeneracies in spectral modelling}\label{subsec:spectral_types}

In Sect.\,\ref{subsec:class}, we demonstrated that the use of different $\beta$-laws can affect the diagnostic line and also the spectral classification.
To better understand this, we show the emission line formation strength $\xi$ \citep[defined by][]{Hillier1987} in Fig.\,\ref{fig:hillier_emission}. 
We see that for different $\beta$-values, the line formation regions occur at different radii and velocities, also corresponding to different local electron temperatures. 
For increasing $\beta$-values, the line formation regions will in general occur at lower temperatures due to the increasing radii of the line formation regions. 
Consequently, stellar models with the same stellar temperature $T_\ast$ but a higher $\beta$-law will appear cooler as the diagnostic lines originate from cooler line formation regions. 
This causes the spectral types of WR-star models to shift in general to a later-type class when the $\beta$-value is increased while keeping the stellar parameters constant.
Naturally, this effect enhances when more extreme values of $\beta$ are adopted (e.g.\ $\beta=20$ or higher).
As a result, the entire range of spectral subclasses can be covered by changing $\beta$ while the model stellar temperature is left unchanged. 
This effect is so extreme for WR stars due to the line spectrum being formed largely or even completely in the stellar wind.

\begin{figure}
    \centering
    \includegraphics[width=\hsize]{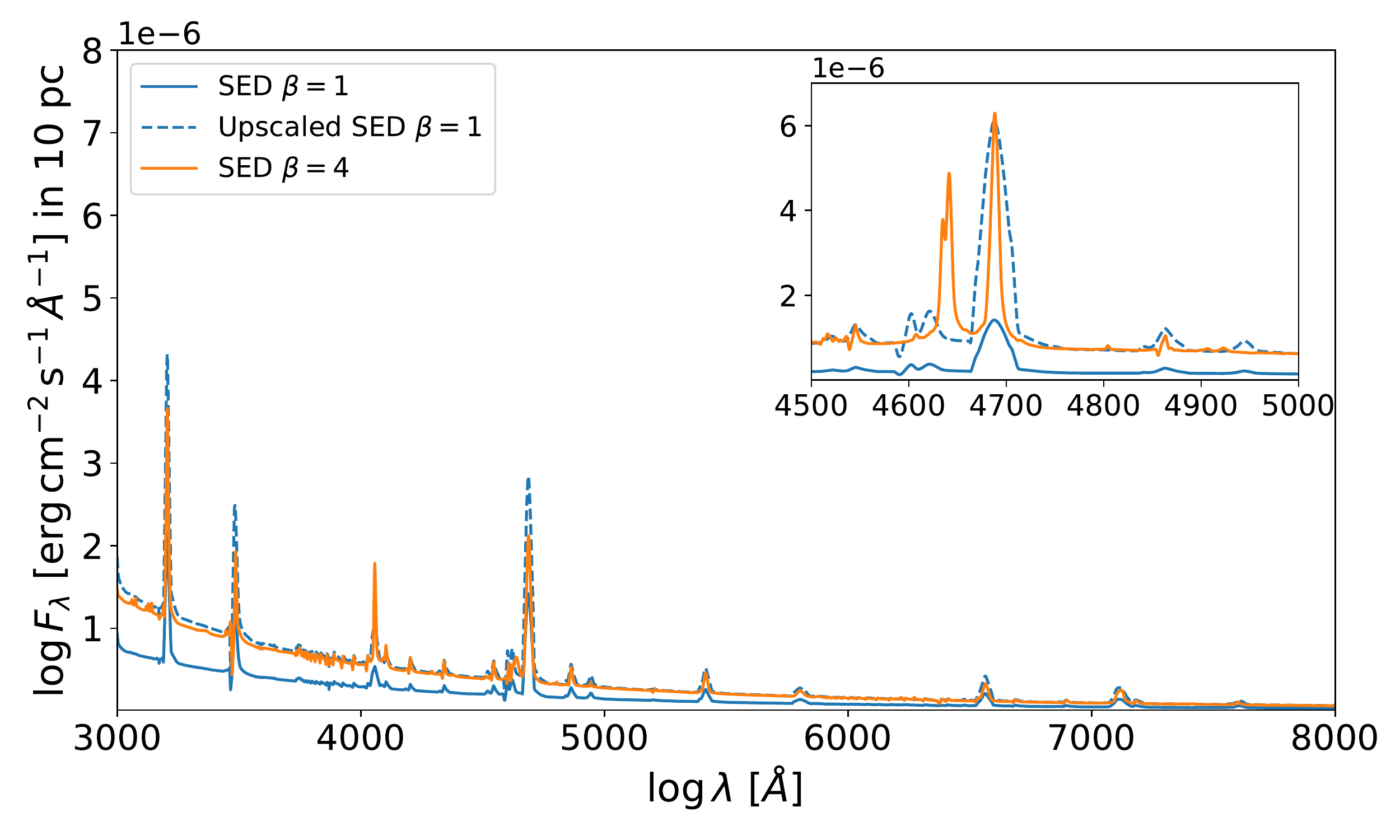}
    \caption{The optical regime of the full SED of $\beta1$ (blue) and $\beta20$ (orange) B2 WN-star models.
    The dashed blue line represents the $\beta1$ flux rescaled (by constant factor 4.3) to the $\beta20$ model flux.
    The inset shows a focus on some of the WN-star diagnostic lines used above (see Tab. \ref{tab:diagnostic_lines}).}
    \label{fig:sed_opt}
\end{figure}

The model continuum is affected by different velocity laws as well (cf.\ Sect.\,\ref{subsec:continuum}).
The continuum peaks shift to longer wavelengths, implying a cooler $T_{2/3}$. 
Due to the different fluxes of the continua, an additional ambiguity can arise in estimating the luminosity. 
When for example reproducing optical photometry with the B2 WN models, a factor of 4.3 is required to match the $\beta1$ flux to the $\beta20$ flux. 
This is illustrated in Fig.\,\ref{fig:sed_opt}, where we look at the optical part of the flux-calibrated spectrum of both models. 
Taking the factor of 4.3 at face value, this would correspond to a significant difference of 0.6\,dex in the derived luminosity. 
As we did not obtain good spectral fits with extreme values of $\beta$,
the actual impact will usually be smaller, but can still be on the order of 0.3\,dex (cf.\ Sect.\,\ref{subsec:obs}).

The obtained differences in the continuum further raise the question whether our derived changes in the normalized line spectrum are intrinsic to the lines or arise from different continuum fluxes. 
The inset in Fig.\,\ref{fig:sed_opt} focuses on a few diagnostic lines in the optical spectrum for two different flux-calibrated models. 
For comparison we further show the $\beta1$ model scaled to the level of the $\beta20$ flux. 
It is evident, that the lines are not diminished by the stronger continuum, but intrinsically change their strength and shape. 
Given that we see similar effects as in the normalised spectra (see, e.g., Fig.\,\ref{fig:opt_line}), we conclude that most of the line changes originate due to the different velocity fields and cannot solely be attributed to normalisation effects.
A study by \citet{ignace2009} was able to show that continuum changes originating from wind structure differences can influence line widths in hot and dense stellar winds. 
Due to the different $\beta$-law assumptions, it is hard to exclude the possibility of additional continuum effects in our work.

We demonstrated previously that -- at least to some degree -- the changes to the spectrum due to a higher $\beta$-value can be mitigated by changing other stellar and wind parameters of the model, causing a degeneracy in the solutions to an observed spectrum.
This degeneracy acts on top of the already known degeneracy in the regime of very dense winds \cite[see, e.g.,][]{hamann2004grids}. 
There, the same optical spectra (from models with fixed $\beta = 1$) can be obtained for different choices of $R_\ast$ (or correspondingly $T_\ast$) if $\log ( R_\text{t}\,[R_\odot]) \leq 0.4$.
Our findings for the different $\beta$-values now yields an additional parameter degeneracy on top of this effect.

\subsection{Implications for the WR radius problem}\label{subsec:radius_problem}

Via the mass continuity equation $\dot{M} = 4\pi r^2 \rho(r)\varv(r)$, we can directly connect the local wind density $\rho(r)$ and velocity $\varv(r)$.
For the standard assumption of a constant mass-loss rate $\dot{M}$, lower velocities directly imply higher wind densities. This causes wind models with $\beta>1$ velocity laws to have a more extended higher-density wind region. 
As a result, the optically thick region of a WR-star wind grows larger and hence the apparent ``photospheric'' radius, i.e.\ $R_{2/3}$, does as well.

This radius-dependency raises the question whether the application of higher $\beta$-values for the velocity fields in the empirical spectral analysis of WR stars may provide a solution to the WR radius problem, see also Sect. \ref{sec:intro}.
Assuming that in general WR stars have winds that could be modelled better using $\beta>1$, the stellar temperatures derived with $\beta=1$ would be underestimated. 
This would shift the observations in Fig.\,\ref{fig:comp} closer to the He-ZAMS. 
However, to completely resolve the large temperature discrepancies between the He-ZAMS and the stellar temperatures derived in the literature, rather extreme values of $\beta$ are required. 
For example, the temperature discrepancy between the $\beta20$ mock observation and the best-fit $\beta1$-model in Fig. \ref{fig:mimick} adds up to $26.1$ kK, causing a considerable temperature shift in the HR-diagram. 
Still, even with a $\beta20$ law the model does not reach the He-ZAMS. 
Hence, to shift the coolest classical WR stars to stellar temperatures close to the He-ZAMS, either more extreme $\beta$-values would be required or the $\beta$-law description is generally insufficient to achieve this goal. 
In Sect.\,\ref{subsec:obs} and Appendix \ref{sec:appendix} we show that extreme values of $\beta > 20$ do not seem to be consistent with observations.
This point will be discussed further in Sect.\,\ref{sec:hydrocmp}.

\section{Comparison with hydrodynamically consistent models}\label{sec:hydrocmp}

In the previous sections, we demonstrated that $\beta$-type velocity laws are able to yield a reasonable reproduction of the observed spectrum, but do not yield parameters for the deeper, hydrostatic layers that are in line with evolutionary expectations. 
Dynamically consistent atmosphere calculations \citep[][]{grafener2005hydrodynamic,sander2020driving} overcome this discrepancy by obtaining $\varv(r)$ directly from solving the hydrodynamic equation of motion  
\begin{align}\label{eq:hydro}
    \varv\left(1 - \frac{a^2}{\varv^2}\right)\frac{d\varv}{dr} & = a_\mathrm{rad} - g + 2\frac{a^2}{r} - \frac{da^2}{dr}\\ 
    & = \frac{GM}{r^2}(\Gamma_\mathrm{rad} - 1) + 2\frac{a^2}{r} - \frac{da^2}{dr},
\end{align}

\noindent with $a$ denoting the isothermal sound speed, $g$ the gravitational acceleration, $a_\mathrm{rad}$ the radiative acceleration ,and $\Gamma_\mathrm{rad} := a_\mathrm{rad}(r) / g(r)$. 
From obtaining $\varv(r)$, $\varv_\infty$ is automatically known. 
Moreover, $\dot{M}$, or alternatively $M_\ast$, have to be iteratively adjusted to obtain a consistent solution.
We note that the dynamical consistency is reached in the stationary case.

\begin{figure}
    \centering
    \begin{subfigure}{\hsize}
        \includegraphics[width=\hsize]{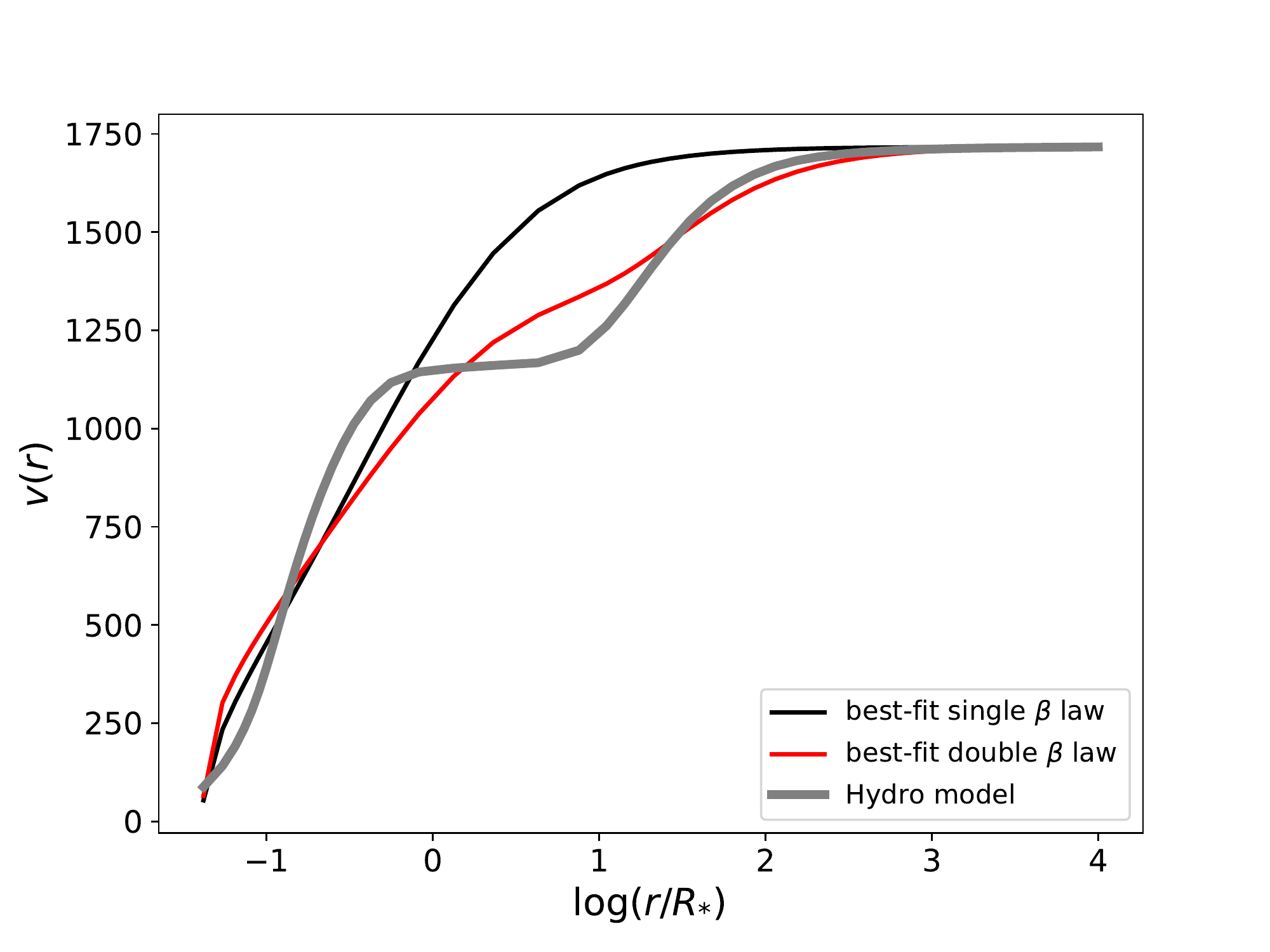}    
    \end{subfigure}\vspace{-7mm}
    \begin{subfigure}{\hsize}
        \includegraphics[width=\hsize]{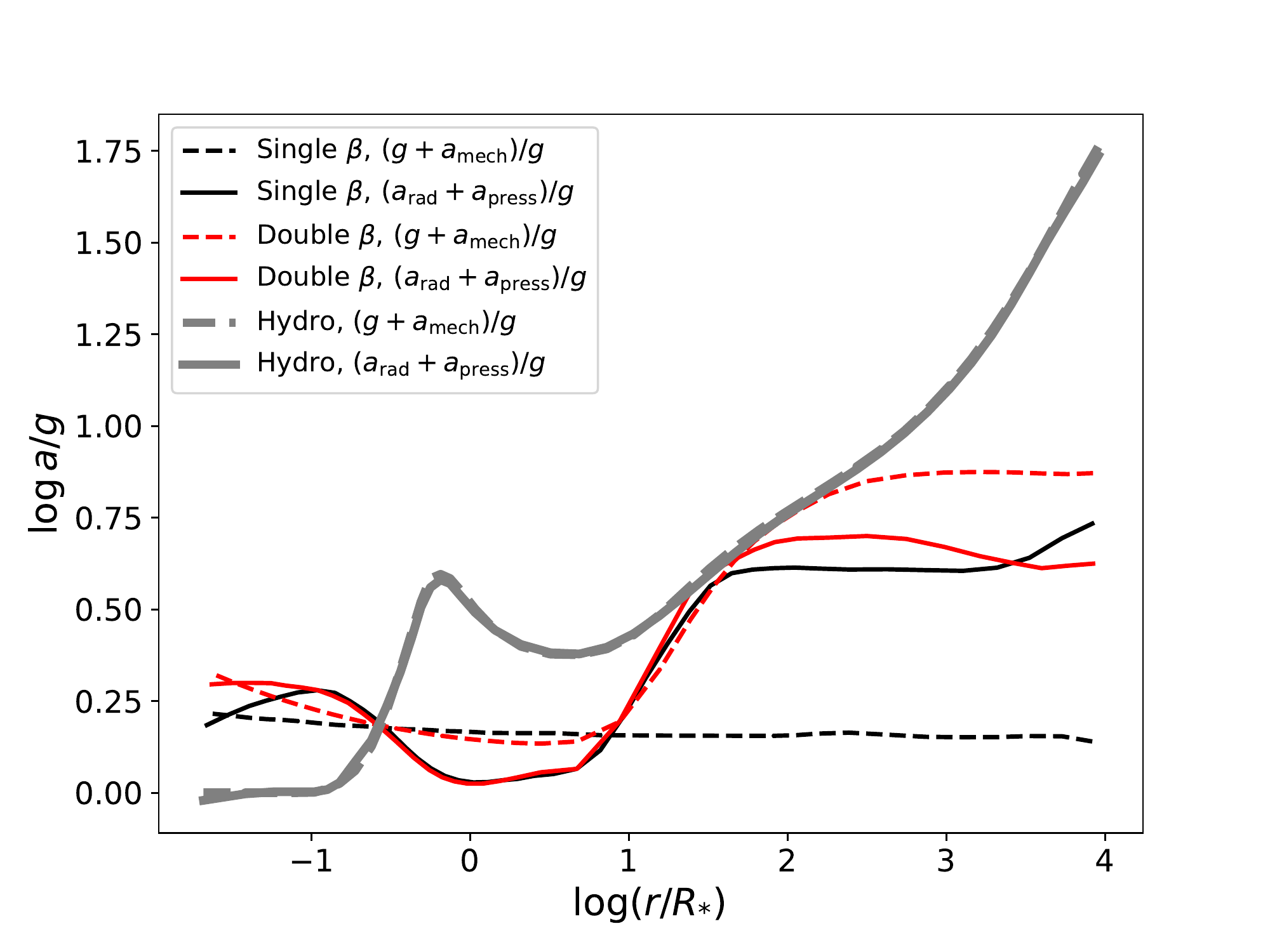}   
    \end{subfigure}
    \caption{Top panel: velocity field of a hydrodynamically consistent WN-star model (bold grey) with best-fit single-$\beta$ (black) and double-$\beta$ (red) velocity laws. 
    Bottom panel: accompanying wind acceleration per model in the same colours.  
    The main deceleration terms ($g$ and $a_\mathrm{mech}$) are dashed, the main acceleration terms ($a_\mathrm{rad}$ and $a_\mathrm{press}$) are the full lines.}
    \label{fig:vlaw_acc_comp}
\end{figure}

The velocity structure of such a hydrodynamically consistent model may look significantly different from the typically assumed single $\beta$ or double-$\beta$ laws. 
To illustrate that, we take a consistent WN-star model from \citet{sander2020nature}, having main parameters $T_\ast = 141$ kK, $\dot{M}_\mathrm{t} = -4.20$, $\varv_\infty = 1718$ km\,s$^{-1}$.
A fit, via $\chi^2$-minimisation, is performed to the resulting $\varv(r)$ with both a single- and double-$\beta$-description and then we run atmosphere models incorporating the $\beta$-type laws with $\varv_\infty$ and all other parameters being identical to the hydrodynamic model. 
The velocity fields and resulting wind acceleration stratifications are shown in Figure \ref{fig:vlaw_acc_comp}. 

\begin{figure*}
    \centering
    \includegraphics[width=\hsize]{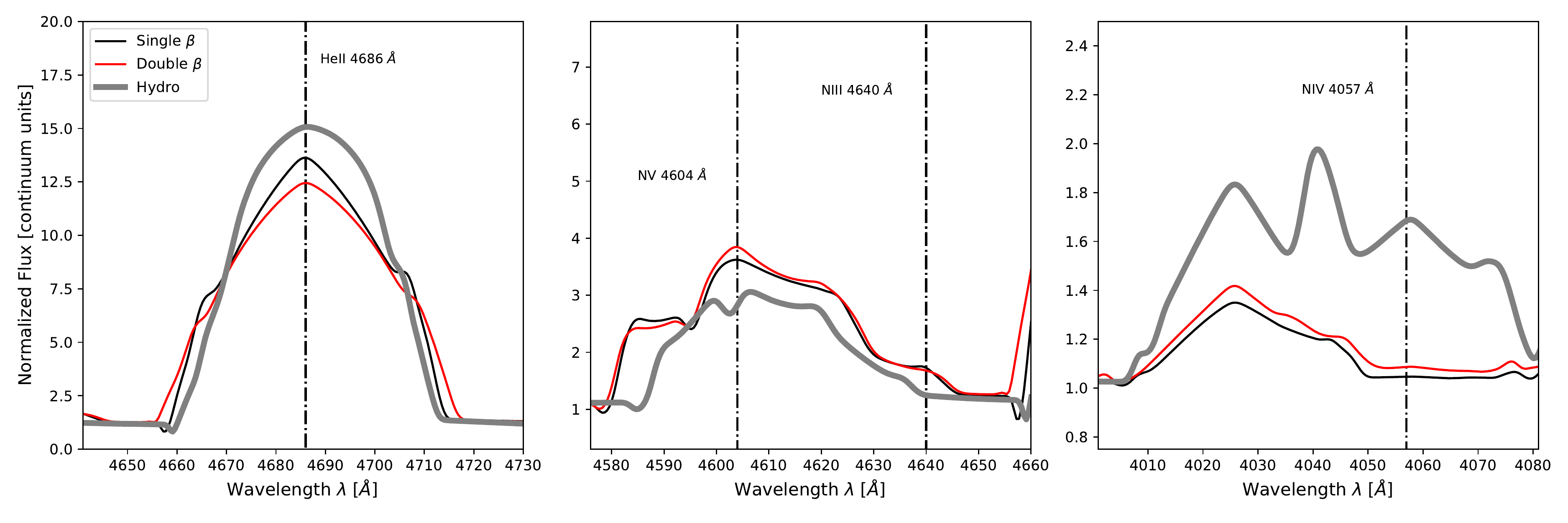}
    \caption{Zoom-in on WN-star diagnostic spectral lines (see also Table \ref{tab:diagnostic_lines}) for the hydrodynamically consistent, single $\beta$ and double $\beta$ WN-star models from Fig. \ref{fig:vlaw_acc_comp}, with the same colours.}
    \label{fig:vlaw_comp_lines}
\end{figure*}

From the panel showing $\varv(r)$, it is immediately evident that the hydrodynamically consistent solution is not well approximated by single- or double-$\beta$-laws.
The wind accelerations show significant differences as well. 
As expected, the hydrodynamically consistent model correctly balances the outward pushing forces (radiation $a_\mathrm{rad}$ and pressure gradient $a_\mathrm{press}$) with the repelling terms (gravity $g$ and inertia $a_\mathrm{mech}$). 
The models using the single and double $\beta$ laws do not only miss the local dynamical consistency, but their overall acceleration terms are also significantly different. 
In particular, the inner acceleration bump caused by the iron M-shell opacities cannot be reproduced. 
This is likely the main reason for the general failure of $\beta$ laws to resolve the radius problem of Wolf-Rayet stars.

Focusing on a few diagnostic lines, the spectral imprint of the different velocity laws are shown in Figure \ref{fig:vlaw_comp_lines}. 
While the single- and double-$\beta$-law models show only minor differences, the hydrodynamically consistent model spectrum deviates strongly from the other two. 
Interestingly, the \ion{He}{ii} 4686\,\AA\ line of the hydrodynamically consistent model in Fig.\,\ref{fig:vlaw_comp_lines} has a more ``round'' shape in comparison with the $\beta$ model spectra. 
Round-shaped lines -- also termed ``bowler-hat''-shaped lines --  have been observed in a sub-sample of WR stars \citep[e.g.][]{hamann2006galactic,2014A&A...565A..27H} including the Galactic target WR2 and various LMC WN stars. 
In order to reproduced their shape, high rotational velocities have been adopted in spectral fitting \citep[e.g.][]{2014A&A...562A.118S,2014A&A...565A..27H}. However, these would require a strong co-rotation of the wind, which would require a dedicated mechanism such as powerful magnetic fields \citep{2014A&A...562A.118S}. 
A detailed study of WR2 star by \citet{2019MNRAS.484.5834C} demonstrated that fast rotation and strong magnetic fields to be unlikely, leaving the origin of the line shapes unclear. 
Given the obtained shape of the \ion{He}{ii} 4686\,\AA\ line in our hydrodynamic model, it might just be the structure of $\varv(r)$ itself yielding this peculiar emission line shape. 
However, a much more tailored study for these kind of objects would be required in order to determine why some, but not all early-type WN stars show this phenomenon.

Beside the line shapes, the dynamically-consistent model further has a notably cooler appearance than the $\beta$-law spectra, reflected in a $T_{2/3}$ difference of about 10\,kK. 
This is quite significant given the former result of $T_{2/3}$ being hardly affected when changing $\beta$. 
Since the stellar parameters of the three models are the same, this effect is purely due to the different velocity and density structure of the models. 

\begin{figure}
    \centering
    \includegraphics[width=\hsize]{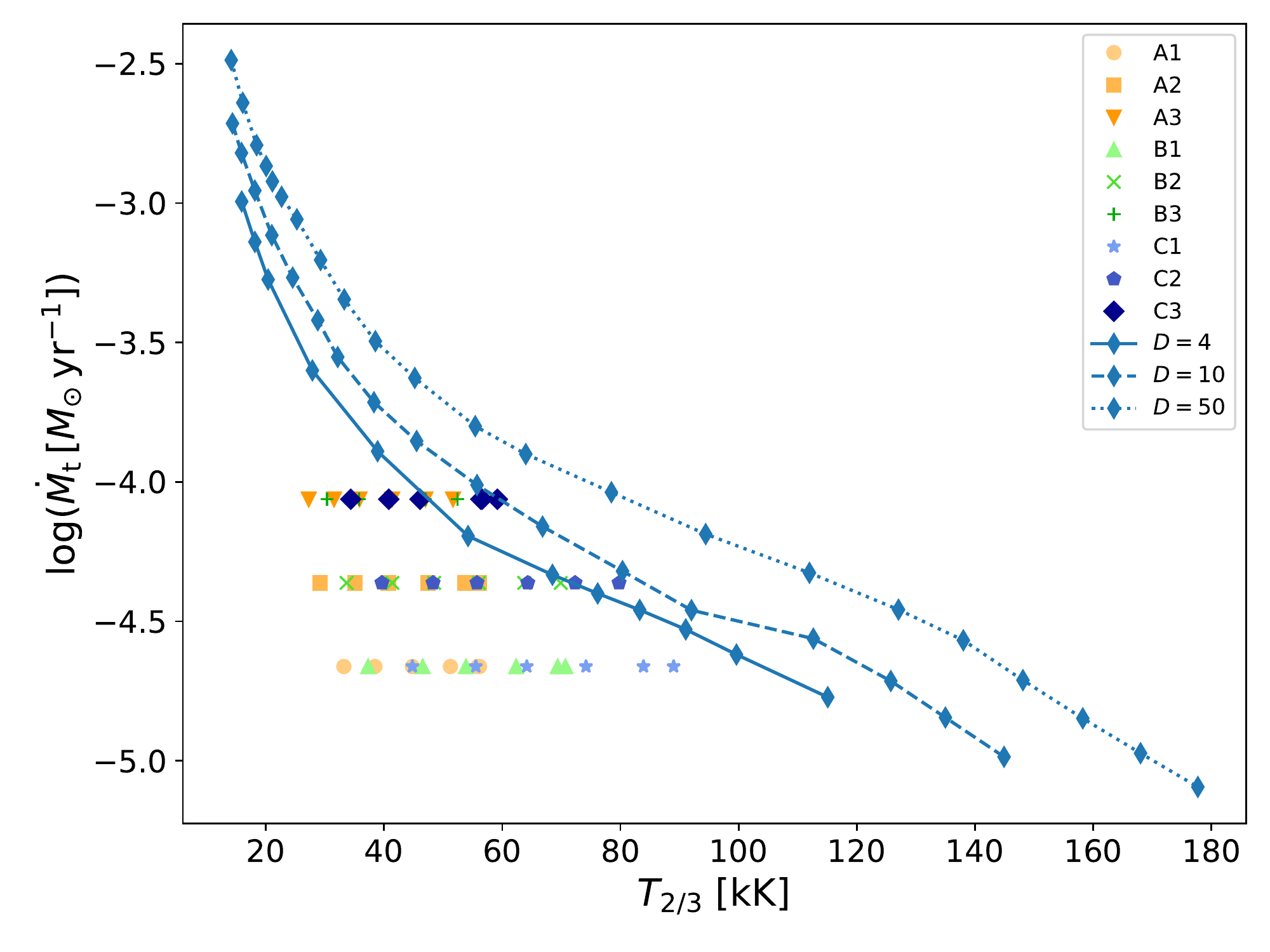}
    \caption{Comparison between hydrodynamically consistent WN-star model sequences from \citet{sander2023temperature} (blue) and the WN-star $\beta$-law models from this study.
    The three hydrodynamic sequences represent different clumping descriptions, while the $\beta$ models are indicated by the same symbols and colours as in Fig. \ref{fig:luka_classif_all}.}
    \label{fig:mdottrans_t23}
\end{figure}

After verifying that the values of $T_{2/3}$ are not significantly affected when reducing the set of considered elements from the hydro models down to those in the typical WR grid models and our above $\beta$-sequences, we now present a more general overview about the $T_{2/3}$-differences between $\beta$-law and hydrodynamically consistent models. 
For this, we consider the transformed mass-loss rates $\dot{M}_\mathrm{t}$ and use the recent result from \citet{sander2023temperature} that hydrodynamically consistent models predict a trend in the $\dot{M}_\mathrm{t}$-$T_{2/3}$-plane that is almost independent of chemical composition and only shifts with different clumping assumptions.
In Fig.\,\ref{fig:mdottrans_t23} we compare the $T_{2/3}$ values from our $\beta$-model sequences (see Tab.\,\ref{tab:gridmodel_params}) with the results from the hydrodynamically consistent models. 
Given the similarities of the different hydrodynamic sequences obtained in \citet{sander2023temperature}, we only plot the curves for the $20\,M_\odot$-sequences at $Z_\odot$ and $X_\text{H} = 0.2$. 
The three shown curves differ only in their clumping description, which is depth-dependent with the maximum clumping factor $D$ being denoted in the figure. 
In contrast, the $\beta$ models (A1 to C3) all have a constant clumping factor of $D=10$.

As expected from our results in Sect.\,\ref{sec:results}, the $\beta$-model sequences show consistently decreasing values of $T_{2/3}$ for increasing $\beta$.
Moreover, the $\beta$-models cover a parameter regime with cooler $T_{2/3}$ temperatures than predicted by hydrodynamically consistent models with winds launched by the hot iron bump. 
For higher values of $\dot{M}_\text{t}$ -- i.e.\ more dense winds -- which are denoted by darker shaded colours, the hotter (blue colored labels in Fig. \ref{fig:mdottrans_t23}) models with the lower $\beta$-values are located near or even on the curve from the hydrodynamic models with a similar ($D = 10$) clumping. 
Especially the lower $\beta$ C3 models agree well with the hydrodynamic model curve.
These results indicate that the traditional WR model grids might partially probe a different parameter space than what we obtain with hydrodynamically consistent models, in particular for less dense WR winds with lower $\dot{M}_\mathrm{t}$ (lighter shaded coloured labels in the figure). 
It is currently an open question whether some of the objects that are well analysed with $\beta$-models can be explained in a dynamically-consistent way without invoking additional effects such as an inflated envelope.

\section{Conclusions}\label{sec:conclusions}

In this work, we have investigated how the velocity law impacts the emergent WR-star spectrum by computing model sequences with different $\beta$-velocity laws using the PoWR code. 
We analysed the resulting differences in the diagnostic lines and spectral-type classifications. 
Detailed modelling revealed that the emergent optical spectra are strongly dependent on the velocity field used in the atmosphere model. 
This effect is so profound that theoretically the entire range of spectral subtypes could be covered with a single set of stellar parameters and only varying the $\beta$-parameter. 
Moreover, the continua are considerably affected by the choice of $\beta$ with higher values yielding higher flux levels in the optical and near-IR regime. 
Below the \ion{He}{i} ionization edge, the trend is reversed with lower $\beta$-models providing higher numbers of \ion{He}{i} ionizing photons. 
There are also effects on the $\ion{H}{i}$ ionizing flux, which decreases more moderately with increasing $\beta$, and the produced Lyman-Werner photons, which tend to increase with increasing $\beta$.

To estimate the actual impact of different $\beta$-laws in spectral analysis, we performed an exemplary fitting of observations in three different environments with $\beta\neq 1$ models and compared to the $\beta1$-model fits. 
Our findings limit the range of reasonable $\beta$-values to $\beta \approx 4$ when restricting the analysis to single-$\beta$ descriptions. 
While this limits the overall impact, we still find a notable ambiguity in determining stellar parameters when relying on $\beta$-type velocity laws, especially concerning the stellar temperature inferred for the hydrostatic layers $T_\ast$. 
The $T_{2/3}$-temperature is less affected within the framework of $\beta$-type laws, but does shift notably when considering a dynamically-consistent velocity law. 
Within the constrained range of $\beta$ values, the derived luminosity is only mildly affected by changing the velocity law. 
Tests with dynamically-consistent models beyond \citet{grafener2005hydrodynamic} will be required to check if this also holds for velocity descriptions beyond the $\beta$ law

When only relying on optical (or infrared) spectra, the terminal velocity $\varv_\infty$ can be severely underestimated when considering only models with $\beta = 1$. 
This problem does not exist if UV spectroscopy with saturated P-Cygni profiles are available, underlining that despite their emission-line spectra, UV spectroscopy is crucial to fully determine the wind (and feedback) parameters of a WR star.
Alternatively, WR-star terminal wind velocities can also be derived from forbidden emission lines \citep[see, e.g.,][]{dessart2000forbidden, ignace2007forbidden}. 
However, these lines are located in the mid-IR and thus IR space telescopes (e.g., JWST MIRI) are required.

By comparing our data from the $\beta$-model sequences to results from hydrodynamically consistent atmosphere modelling, we show that $\beta$-type velocity laws are insufficient to describe the inner wind regime. 
This insufficiency also remains when using a double-$\beta$ approach tailored to the hydrodynamic velocity solution.
Given the fact that empirical analyses with models using $\beta \lesssim 4$ provide good spectral reproductions, the outer regions of WR winds seem to be reasonably well described by a single $\beta$-law, despite the shortcomings in the inner wind.
From a first test case where we compare a dynamically-consistent solution to the ``best'' $\beta$-approximations, we conclude that the spectral line shapes are notably affected when trying to approximate with a $\beta$ or double-$\beta$ law. 
Interestingly, the emission lines are more roundish in the hydrodynamically consistent model, hinting at a possible origin for the special ``round''/``bowler-hat'' line shapes found for WR2 and various LMC stars.

In a follow-up study, we will investigate how to improve the description of WR-star velocity fields by performing detailed analyses of individual WR stars in different environments with hydrodynamically consistent atmosphere models.

\section*{Acknowledgements}

This research is based on observations made with the NASA/ESA Hubble Space Telescope obtained from the Space Telescope Science Institute, which is operated by the Association of Universities for Research in Astronomy, Inc., under NASA contract NAS 5–26555.
These observations are associated with program 15822 (PI A.A.C.\ Sander).
This work is funded by the Deutsche Forschungsgemeinschaft (DFG, German Research Foundation) -- Project-ID 138713538 -- SFB 881 (``The Milky Way System'', subproject P04). 
AACS and RRL are supported by the Deutsche Forschungsgemeinschaft (DFG, German Research Foundation) in the form of an Emmy Noether Research Group -- Project-ID 445674056 (SA4064/1-1, PI Sander)
TS acknowledges support from the European Union’s Horizon 2020 under the Marie Skłodowska-Curie grant agreement No 101024605.
LP acknowledges support from the KU Leuven C1 grant MAESTRO C16/17/007.
KD acknowledges funding from the European Research Council under European Unions Horizon 2020 research programme (grant agreement No. 772225).
We also want to thank the referee for their helpful comments and suggestions.

\section*{Data availability}

The data underlying this article will be shared on reasonable request to the corresponding author.

\bibliographystyle{mnras}
\bibliography{biblio}

\appendix

\section{Spectral fits}\label{sec:appendix}

The spectral fitting results from Sect. \ref{subsec:obs} are detailed here, giving a special emphasis on optical diagnostic lines and the \ion{C}{iv} $\lambda\lambda$ 1548-50 P-Cygni profile in the UV.
See also Table \ref{tab:obsdat_stars} for the resulting parameters.

\begin{figure*}
    \centering
    \includegraphics[width=0.97\hsize]{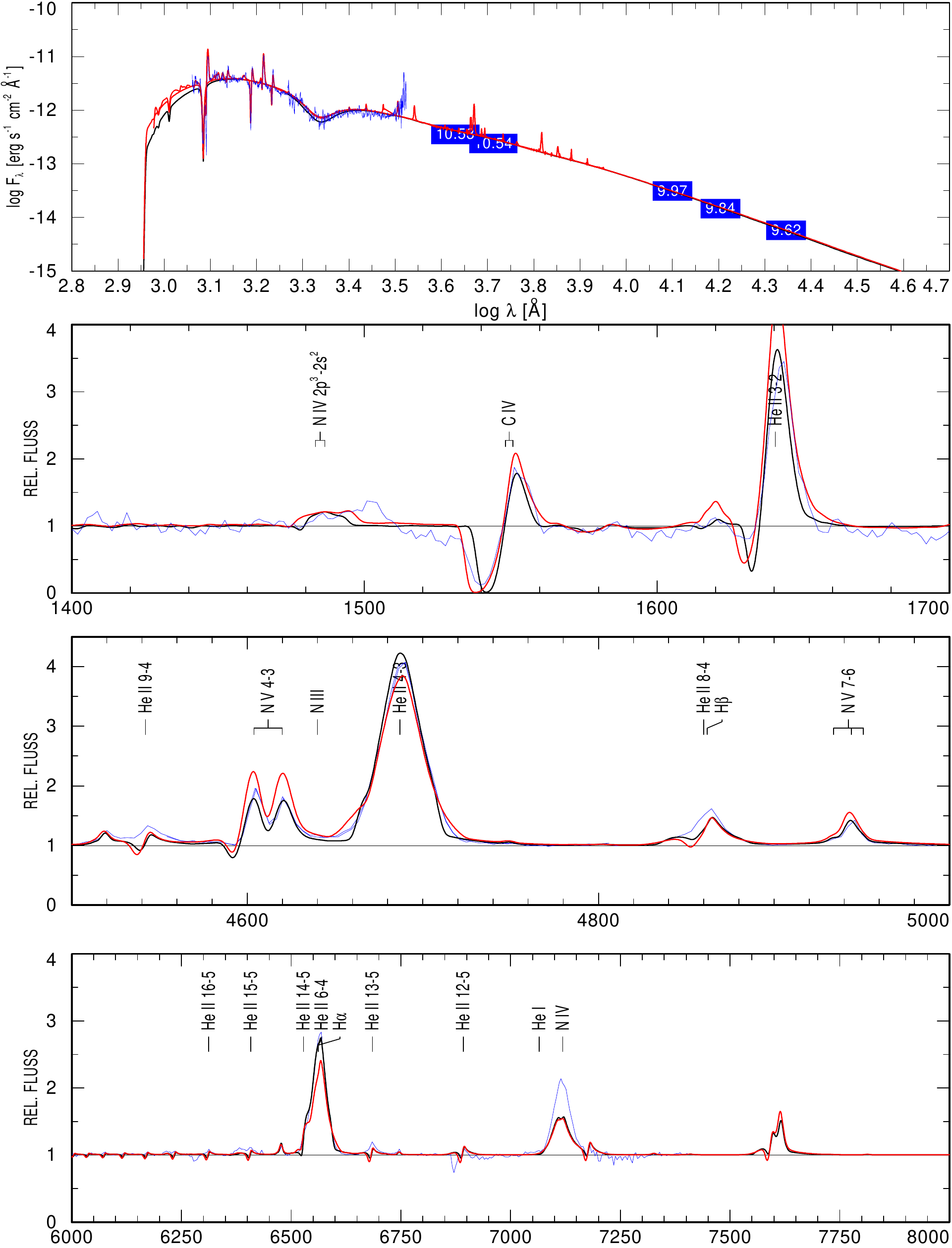}
    \caption{Observed (blue) and model (black and red) spectra of WR128. 
    The black model spectrum was computed using a $\beta1$-velocity law, while the red model spectrum was computed with $\beta=4$.
    The $\beta1$ model continuum does not display the spectral lines in the top plot.}
    \label{fig:wr128}
\end{figure*}

\paragraph*{WR128:} The spectral data from \citet{1995A&AS..113..459H} are used for this star, which is classified as WN4(h)-w \citep{hamann2019galactic}. 
The resulting fit is shown in Fig.\,\ref{fig:wr128}, where $\beta1$-and $\beta4$-models are compared to the observed spectrum. 
Higher $\beta$-values do not longer provide a reasonable representation of the observed spectrum. 
There are obvious deficiencies between the two models and the observed spectrum in Fig.\,\ref{fig:wr128}, but on a level which would e.g.\ be acceptable in a larger sample analysis with pre-calculated grids. 
In such a context, both models provide a reasonable representation of the optical spectrum (cf.\ the bottom two panels).

The corresponding model parameters are shown in Table\,\ref{tab:obsdat_stars}, illustrating significant differences between the models. 
We see a $T_\ast$ difference of $\Delta T_\ast = 18.4$ kK, $\Delta \varv_\infty = 800$ km\,s$^{-1}$ and $\Delta \log\dot{M} = 0.2$ (amounting to about $60$\%) between the new $\beta1$ and the $\beta4$ model. 
This example underlines our previously obtained ambiguity when using different $\beta$-velocity laws: a good agreement to the optical spectrum can be found with significantly different model parameters. 
When considering also the UV (second panel), none of the models fully reproduce the shape of the \ion{C}{iv} $\lambda\lambda$ 1548-50 line.
Still, it is clear that the absorption trough is broader than reproduced by the $\beta1$-model. 
The $\beta4$-model, which also requires a higher $\varv_\infty$ to reproduce the optical part of the spectrum, provides a better representation of the wind velocity range here.
The continua of both models agree fairly well, where there are some differences in the reddening and derived luminosities (see Tab.\ref{tab:obsdat_stars}).

\begin{figure*}
    \centering
    \includegraphics[width=0.97\hsize]{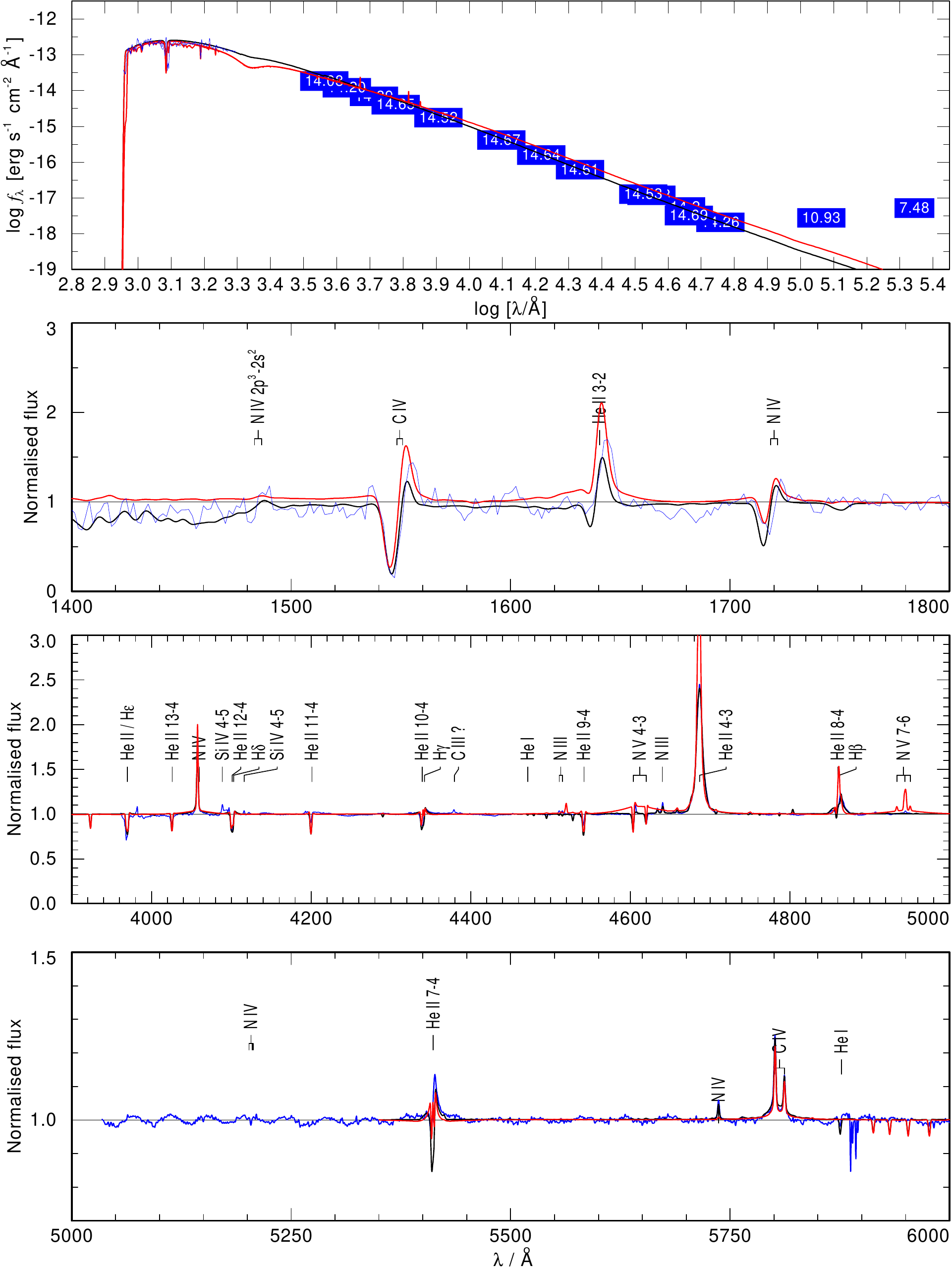}
    \caption{Observed (blue) and model (black and red) spectra of SMC AB2. 
    The black model is the original fit from \citet{Hainich+2015} where $\beta=1$.
    The red model was computed with $\beta=4$.
    The $\beta1$ model continuum does not display the spectral lines in the top plot.}
    \label{fig:smcab2}
\end{figure*}

\paragraph*{SMC\,AB2:} This star is classified as WN5ha in \citet{Hainich+2015} and has an even weaker wind than WR\,128. 
The spectral fitting to the observed spectrum is shown in Fig.\,\ref{fig:smcab2}, displaying the $\beta1$-model from \citet{Hainich+2015} along with the observation and a new $\beta4$-model.
Both models fit the optical spectrum (bottom two panels) quite well, with the $\beta4$-model overestimating some emission line strengths when compared to the $\beta1$-model. 
The UV spectrum (second panel) is also decently reproduced by both models. 
The continua (top panel) agree less well however, with significant differences arising in the mid and far IR regime for the $\beta4$ model.
There is again a significant difference in the model parameters (see Table \ref{tab:obsdat_stars}) with especially $\Delta \log\dot{M} = 1.0$. 
It should be noted that the $\beta4$-model requires an unlikely high clumping factor of $D=130$, making the $\beta1$-model more plausible here, especially when considering the deviating continuum of the $\beta4$ model.
Additionally, the P-Cygni trough of e.g.\ the \ion{C}{iv} $\lambda\lambda$ 1548-50 doublet is reproduced well by the $\beta1$-model. 
We thus conclude that in this case $\beta = 1$ remains the best approximation when focusing on the reproduction of the emergent spectrum. 
In a previous analysis of SMC\,AB2, \citet{Martins+2009} also used $\beta = 1$ and argued that it was best suited to reproduce the shape of the of optical \ion{He}{ii} $\lambda$ 4686 emission line. 
When considering both WR128 and SMC\,AB2, the line seems to indeed have some indicative power, but as we will see in the next example, this might be limited to this particular regime of weaker-wind WN stars if they are of intermediate subtype. 

\begin{figure*}
    \centering
    \includegraphics[width=0.97\hsize]{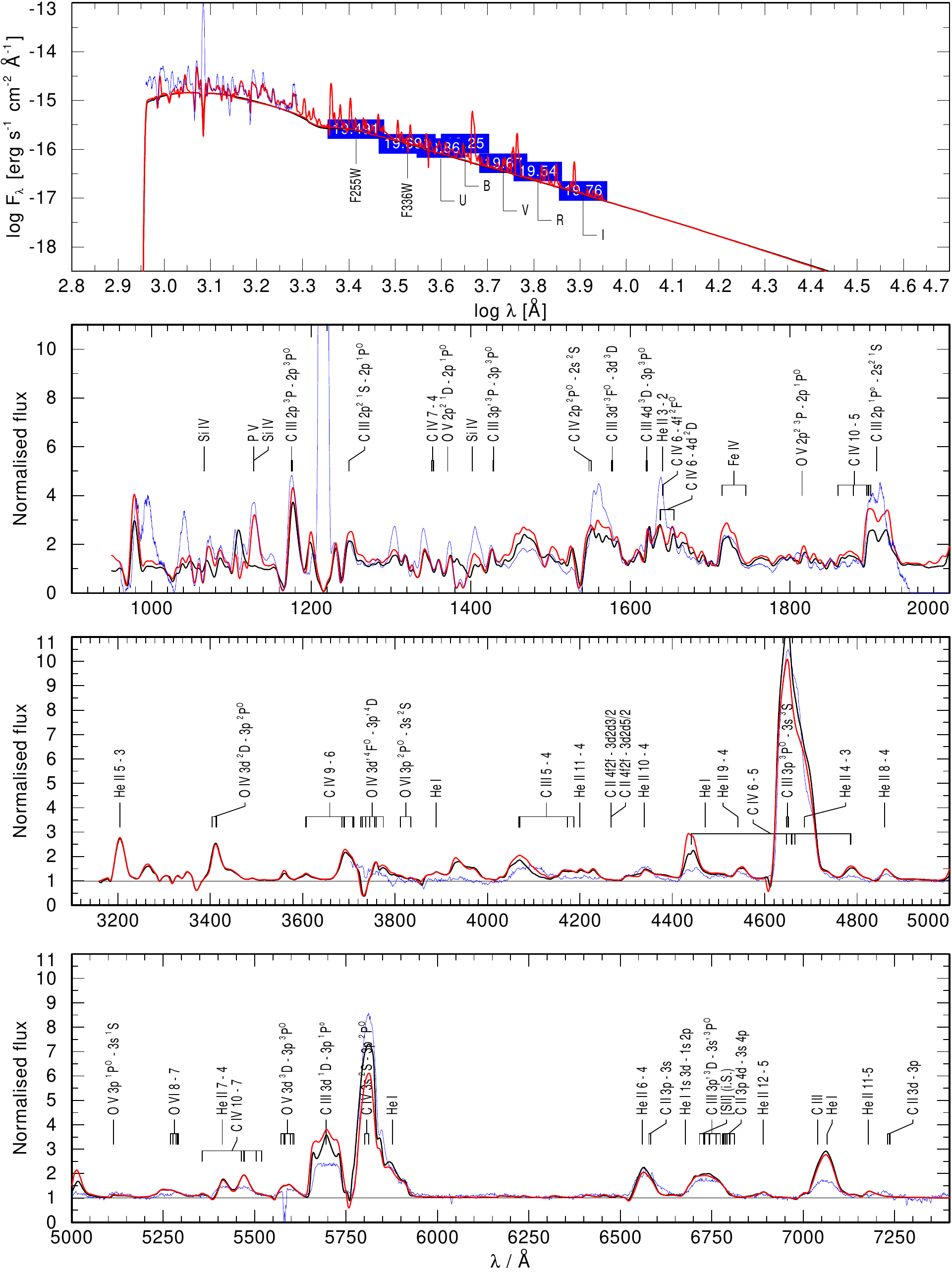}
    \caption{Observed (blue) and model (black and red) spectra of M31\,WR096 (LGGS J004412.44+412941.7). 
    The black model is the $\beta1$-model. 
    The red model was computed with $\beta=4$.
    The $\beta1$ model continuum does not display the spectral lines in the top plot.}
    \label{fig:m31_wr96}
\end{figure*}

\paragraph*{M31\,WR096 (LGGS\,J004412.44+412941.7):} Having a completely different spectral appearance -- classified as WC6 in \citet{Neugent+2012} -- than the former two, we pick a WC-type target from M31, applying the naming scheme introduced in \citet{Sander+2014}. 
Novel UV spectra of this target were observed with the HST, program 16170, and are shown in Fig.\,\ref{fig:m31_wr96} together with a reasonable fit with both a $\beta1$ and a $\beta4$-model.
Despite some discrepancies between the observed and the model spectra, most diagnostic lines (Table \ref{tab:diagnostic_lines}) are reproduced relatively well. 
Similar to the two former observed targets, there is a stellar parameter discrepancy between both models (see Table \ref{tab:obsdat_stars}), most notably in $T_\ast$ where $\Delta T_\ast = 17$ kK.
When only considering the optical spectrum in the bottom two panels, the two model spectra are mostly identical with some difference arising from slightly different mismatches for a few particular lines. 
The model ambiguity also largely continues in the UV-spectrum (second panel), although one could argue that the \ion{C}{iv} $\lambda\lambda$ 1548-50 absorption trough is slightly better reproduced by the $\beta4$-model.
Both continua of the two models are in good agreement as well, with only minor differences in reddening and luminosity.

\begin{figure*}
    \centering
    \includegraphics[width=\hsize]{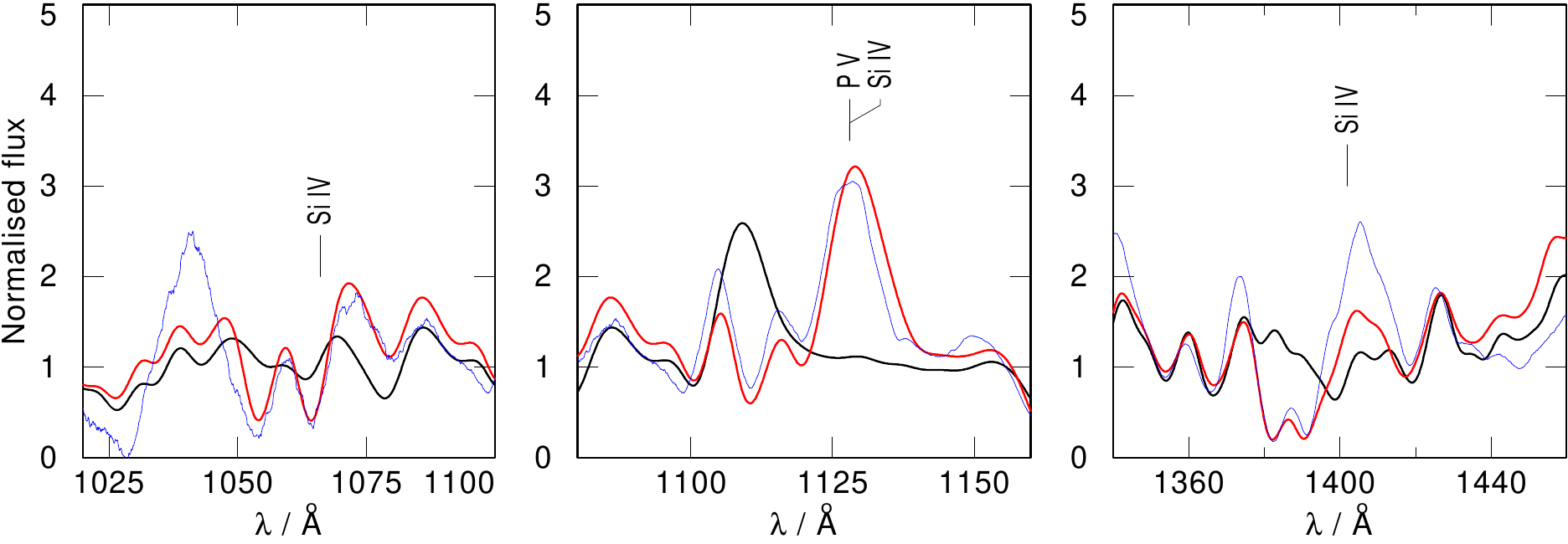}
    \caption{Selection of prominent emission lines from the emergent spectrum of the M31\,WR096 $\beta4$-model. 
    The black spectrum represents the model without Si, P, S and Ne, while the red spectrum includes those elements. 
    While there are no clear signatures of S and Ne-presence in the emergent spectra, the Si and P-signatures shown are prominent.}
    \label{fig:m31wr96elements}
\end{figure*}

Given that M31\,WR096 has never been analysed with atmosphere models before, we also include silicon, phosphorus, sulfur, and neon in our models assuming solar abundances. 
The latter two (S and Ne) do not have noticeable spectral lines in our available observations, but affect the strength of other lines and improve the overall fit quality. 
The first two elements (Si, P) have clear wind lines in the UV, to be seen in Fig.\,\ref{fig:m31wr96elements}. 
However, a further detailed abundance study is beyond the scope of the present paper.
The model spectra in Fig. \ref{fig:m31_wr96} both contain these extra elements.

\bsp	
\label{lastpage}
\end{document}